\documentclass{article}

\usepackage{PRIMEarxiv}

\usepackage[utf8]{inputenc} 
\usepackage[T1]{fontenc}    
\usepackage{hyperref}       
\usepackage{url}            
\usepackage{booktabs}       
\usepackage{amsfonts}       
\usepackage{nicefrac}       
\usepackage{microtype}      
\usepackage{lipsum}
\usepackage{fancyhdr}       
\usepackage{graphicx}       
\usepackage{natbib}    
\usepackage{multirow}

\usepackage{amssymb}
\usepackage[figuresright]{rotating}
\usepackage{subfigure}
\usepackage{amsmath}
\usepackage{hyperref}
\usepackage{enumerate}
\usepackage{color}
\usepackage{tabu}
\usepackage{xr}
\usepackage{threeparttable}
\usepackage{chngcntr}
\usepackage{setspace}
\usepackage{mdwlist}
\usepackage{diagbox}
\usepackage{rotating}
\usepackage{float}
\usepackage{setspace}
\usepackage{comment}
\usepackage[section]{placeins}
\graphicspath{{media/}}     

\title{Covariate-Guided Bayesian Mixture of Spline
Experts for the Analysis of Multivariate Time
Series
}

\author{
  Haoyi Fu \\
  Department of Biostatistics \\
  University of Pittsburgh \\
  Pittsburgh, PA, USA\\
  \texttt{haf48@pitt.edu} \\
   \And
  Lu Tang \\
  Department of Biostatistics \\
  University of Pittsburgh \\
  Pittsburgh, PA, USA \\
  \AND
  Ori Rosen \\
  Department of Mathematical Sciences \\
  University of Texas at El Paso \\
  El Paso, TX, USA \\
  \And
  Alison E. Hipwell \\
  Department of Psychiatry \\
  University of Pittsburgh \\
  Pittsburgh, PA, USA \\
  \And
  Theodore J. Huppert \\
  Department of Electrical and Computer Engineering \\
  University of Pittsburgh \\
  Pittsburgh, PA, USA \\
  \And
  Robert T. Krafty \\
  Department of Biostatistics and Bioinformatics \\
  Emory University \\
  Atlanta, GA, USA \\
}

\begin{document}
\maketitle

\begin{abstract}
With rapid development of techniques to measure brain activity and structure, statistical methods for analyzing modern brain-imaging play an important role in the advancement of science. Imaging data that measure brain function are usually multivariate time series and are heterogeneous across both imaging sources and subjects, which lead to various statistical and computational challenges. In this paper, we propose a group-based method to cluster a collection of multivariate time series via a Bayesian mixture of smoothing splines. Our method assumes each multivariate time series is a mixture of multiple components with different mixing weights. Time-independent covariates are assumed to be associated with the mixture components and are incorporated via logistic weights of a mixture-of-experts model. We formulate this approach under a fully Bayesian framework using Gibbs sampling where the number of components is selected based on a deviance information criterion. The proposed method is compared to existing methods via simulation studies and is applied to a study on functional near-infrared spectroscopy (fNIRS), which aims to understand infant emotional reactivity and recovery from stress. The results reveal distinct patterns of brain activity, as well as associations between these patterns and selected covariates.
\end{abstract}


\section{Introduction}
\label{sec:intro}

Time series are realizations of random processes. Obtaining estimated time series trajectories may provide insights into many practical problems. Functional near-infrared spectroscopy (fNIRS) is a noninvasive brain imaging technique that measures changes in both oxy- and deoxy-hemoglobin using near-infrared light \citep{jobsis1977noninvasive}. In fNIRS, processed data are nonstationary multivariate time series with a non-constant mean and high variability across time, which pose many statistical challenges in inference and estimation. In the case of fNIRS, different subjects could have distinct patterns of multivariate time series trajectories, which could be associated with certain clinical or demographic characteristics. The analysis of fNIRS data requires an appropriate method for the analysis of a collection of multivariate time series observed from different subjects, which is often referred to as a replicated multivariate time series setting. 

Cluster analysis is often used to address the issue of heterogeneity and identify subgroups from collections of time series observed from different subjects. Time series clustering has been used in diverse scientific areas to discover trajectory patterns, which can uncover valuable information from complex and massive datasets \citep{liao2005clustering}. Time series clustering partitions the entire collection of data into different groups such that homogeneous time series are grouped together based on a certain similarity measure. Challenges in time-series clustering include computational issues due to high-dimensionality and the selection of proper similarity measures \citep{lin2003clustering, keogh2000simple}. Several authors have proposed clustering algorithms for multivariate time series. \citet{kakizawa1998discrimination} used Kullback-Leibler discrimination information as the minimum discrimination criterion for clustering multivariate Gaussian time series. \citet{wang2007structure} used a modified $K$-means clustering algorithm for clustering multivariate time series based on univariate structures. A variety of papers have established different model-based clustering methods for clustering multivariate time series, such as multivariate autoregressive models \citep{maharaj1999comparison, he2022arzimm}, a hidden Markov model \citep{li2001building} and smoothing splines \citep{krafty2017conditional,li2019adaptive}. Comprehensive review of methods for time series clustering can be found in \citet{liao2005clustering} and in \citet{maharaj2019}. 

Covariate-dependent structures can often be associated with the mixture components from a clustering of time series. \citet{bertolacci2022adaptspec}  presented an analysis of multiple nonstationary time series by using a covariate-dependent infinite mixture with logistic stick-breaking weights, where mixing weights are computed based on covariates. The mixture-of-experts model \citep{jacobs1991adaptive} assigns weights to each expert via a covariate-dependent multinomial logists. \citet{huerta2003time} addressed the issue of time series model mixing based on covariates using the hierarchical mixture-of-experts \citep{jordan1994hierarchical}.

Smoothing splines, which are nonparametric methods that utilize roughness-based penalties, have been widely used in the analysis of time series \citep{wang2011smoothing, gu2013smoothing}.  Bayesian interpretations of smoothing splines were first discussed by \citet{kimeldorf1970correspondence}. \citet{wahba1978improper} showed that the solution to the smoothing splines objective function is equivalent to Bayesian estimation with a partially diffuse prior. \citet{speckman2003fully} adopted a fully Bayesian approach for implementing smoothing splines with a noninformative prior on the variance component, as well as derived necessary and sufficient conditions for the propriety of the posterior. Smoothing splines require estimation of a large number of coefficients, which might be impractical in high-dimensional settings. \citet{gu2002penalized} used a subset of reproducing kernel functions to achieve a low-dimensional approximation. \citet{wood2002bayesian} obtained a subset of basis functions using the eigen-decomposition of the Gaussian kernel. \citet{krafty2017conditional} proposed a tensor-product model for the analysis of replicated multivariate time series which decomposes the power spectrum into products of univariate outcomes and frequencies.

Our goal in this paper is to perform a covariate-guided clustering of multivariate time series that can capture trajectory patterns of mixture components and evaluate the relationship between covariates and trajectory patterns. To this end, each mixture component is modeled via smoothing splines, and   time-independent covariates are incorporated into the mixture model via the mixing weights. The method is formulated in a fully Bayesian framework. The rest of this paper is organized as follows. In Section \ref{sec:motivating} we introduce the motivating study. Sections \ref{sec:model} and \ref{sec:prior} present the proposed model and priors. Section \ref{sec:schemeP2} introduces the sampling scheme. In Section \ref{sec:sim} we report simulation results under different settings and Section \ref{sec:realdataapp} illustrates our proposed method with application to the motivating study. Section \ref{sec:discussion} concludes the paper with a discussion.  

\section{Motivating Study}
\label{sec:motivating}

Our motivating study aims to understand patterns of infant's brain activity before, during and after an emotionally stressful probe called face-to-face still-face (FFSF) \citep{tronick1978infant}. Participant mothers in this study were recruited from the longitudinal Pittsburgh Girls Study (PGS), a population-based study of 2,450 girls who were recruited in the city of Pittsburgh between the ages of 5 and 8 \citep{keenan2010pittsburgh}. In 2016, a large-scale sub-study of the PGS was initiated to investigate how environmental factors, such as psychological stressors experienced during childhood and adolescence, affect later maternal pregnancy and child health. The study is part of the National Institutes of Health Environmental Influences of Child Health Outcomes (ECHO) program, which examines different impacts of prenatal environmental exposures across biological, chemical, physical and social domains on offspring health and development \citep{gillman2018environmental}. The PGS-ECHO study enrolls PGS participants as they become pregnant or recently deliver a live birth. Participants complete multiple prenatal lab visits and the children are followed from ages 6 to 36 months. The lab protocol includes interviews and interaction tasks to assess contextual stressors, health, mood, lifestyle behaviors and offspring behavioral and emotional development. 

Face-to-face interactions between mothers and infants are essential to the development of infants with respect to communication and social skills, as well as the regulation of emotion and temperament \citep{hipwell2019transgenerational}. The FFSF paradigm is a widely used stress task (a violation of the expectation of social interaction) that allows for biobehavioral measurement of individual differences in infant response and recovery. The FFSF comprises of three phases: interact (or baseline), still-face and recovery \citep{adamson2003still}. In phase 1, mothers perform normal interactions with infants without the use of toys; this phase serves as the baseline. In phase 2, mothers adopt a neutral facial expression (still-face with no facial or oral communication) to infants, followed by phase 3, where mothers resume normal interactions with their infants. Prior to the start of the FFSF, an fNIRS cap is fitted on the infant's head to measure the level of and change in brain activation across the three phases.

\begin{figure}[!p]
\centering
\includegraphics[width=1.0\textwidth]{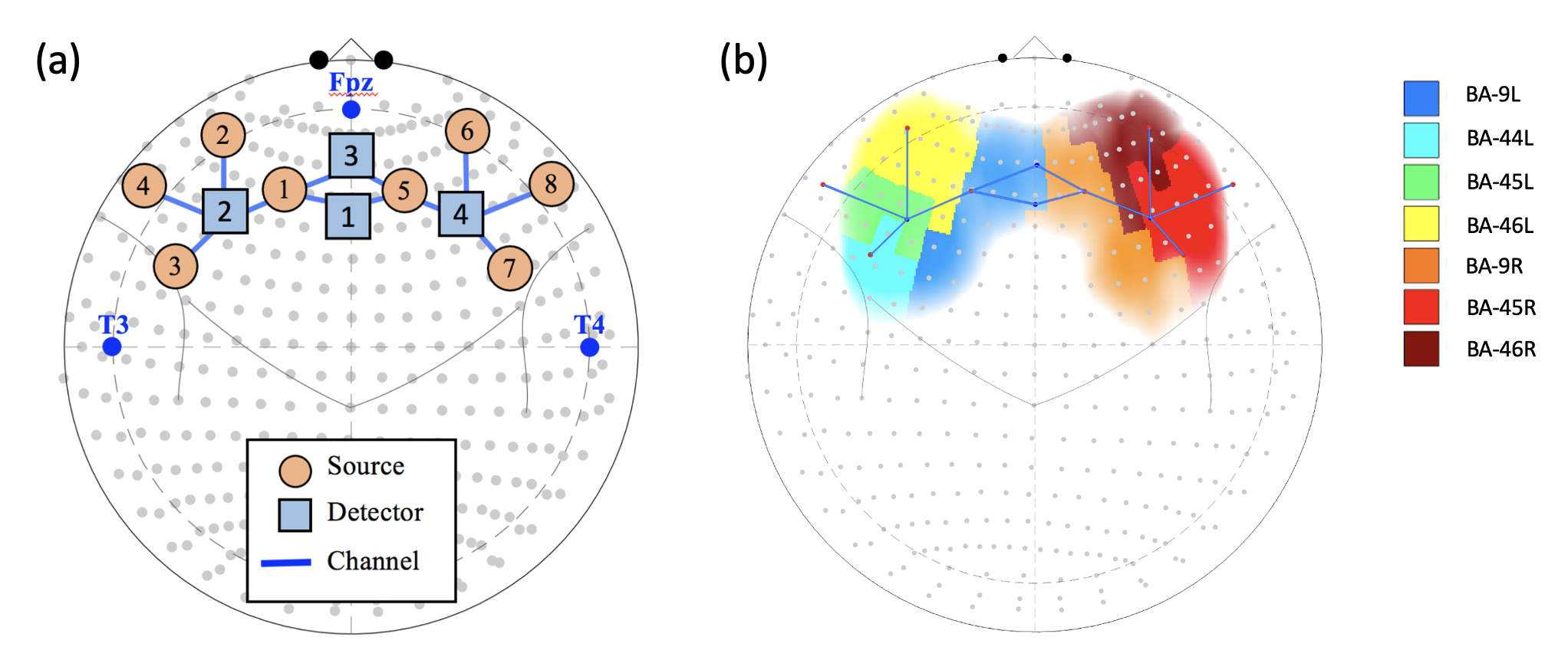}
\caption{fNIRS probe configuration. (a) Positioning of 8 sources, 4 detectors and 12 channels. A channel is connected by one source and one detector (blue line). (b) Brodmann areas covered by fNIRS probe.}
\label{fig:fNIRSprobe}
\end{figure}

PGS-ECHO fNIRS still-face data are recorded using a continuous NIRS imaging system (NIRScout; NIRx Medical Technologies, Berlin, Germany) at the sampling rate of 7.8125 Hz and using the NIRStart acquisition software. The data are measured simultaneously at two wavelengths (760 nm and 850 nm). As shown in Figure \ref{fig:fNIRSprobe}(a), this fNIRS probe consists of 12 channels from 8 sources and 4 detectors.

\begin{figure}[!p]		
	\centering
	\includegraphics[width=1.0\textwidth]{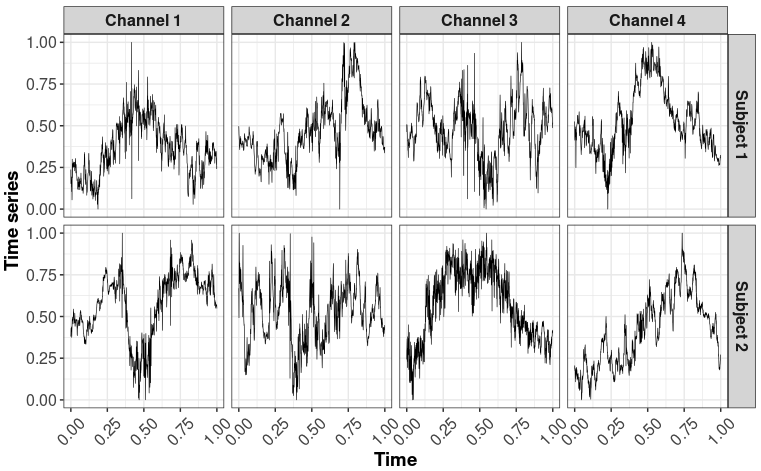}
	\caption{An example of processed fNIRS time series from two selected subjects and four selected channels. The measurements are the relative concentration of oxy-hemoglobin.}
	\label{fig:processedfNIRSts}
\end{figure}

In the current study, we measured infant brain activity using the above fNIRS probe (roughly $120$ seconds of measurements for each phase). At the end of 2021, recorded fNIRS still-face data had been collected from 155 infant subjects. Demographic variables of infants and mothers such as gestational age, infant age, sex, birth weight, head circumference, along with parent reports on the Infant Behavior Questionnaire-Revised (IBQ-R) \citep{gartstein2003studying} were also collected. By removing infants who did not complete the three phases of the still-face paradigm, who had large outliers based on leverage and who had a very short period of measurements in any of the three still-face phases, there were a total of 82 subjects with complete fNIRS still-face data available for future analysis. The above quality control steps were performed by the NIRS brain AnalyzIR toolbox in MATLAB \citep{santosa2018nirs}. Moreover, additional data pre-processing steps were performed in R software, including data interpolation and rescaling. Finally, processed fNIRS data had a total of 1,500 measurement points for each subject and each channel, where each phase consisted of 500 points. All measurements and sampling times were rescaled to be between 0 and 1, with the interact phase occurring between time 0 to 1/3, still-face between 1/3 to 2/3, and recovery between 2/3 to 1. An example of processed fNIRS time series from two selected subjects and four selected channels is displayed in Figure \ref{fig:processedfNIRSts}.

The goals of our analysis are to identify distinct patterns of brain activity trajectories from multiple fNIRS channels represented by the relative concentration of oxy-hemoglobin, and to assess the association between trajectory patterns and relevant covariates. 

\section{Model}
\label{sec:model}

In this section, we provide a detailed description of our proposed covariate-guided Bayesian mixture of spline experts model. The proposed model consists of spline components whose mixing weights depend on covariates.

\subsection{Mixture of splines model}
\label{sec:mixturespline}

We propose a tensor-product mixture of splines model for multivariate time series. For each subject $i = 1, \ldots, N$, let $\boldsymbol{y}_i=(\boldsymbol{y}_{i1}^{\prime},\ldots,\boldsymbol{y}_{ik}^{\prime},\ldots,\boldsymbol{y}_{iK}^{\prime})^{\prime}$ be the $nK$-vector corresponding to the $K$-dimensional time series for $k=1,\ldots,K$, where $\boldsymbol{y}_{ik}=\big[ y_{ik}(t_1),\ldots, y_{ik}(t_j),\ldots, y_{ik}(t_n) \big]^{\prime}$ contains the trajectory of measurements on the $k$th entry of the time series evaluated over a grid of $n$ time points for $j=1,\ldots,n$, and $\boldsymbol{\epsilon}_i= (\boldsymbol{\epsilon}_{i1}^{\prime},\ldots,\boldsymbol{\epsilon}_{iK}^{\prime})^{\prime}$ is the $nK$-vector of errors. Following the model representation of \citet{krafty2017conditional}, the tensor-product model for the $K$-dimensional multivariate time series, conditional on component $g$, $g=1,\ldots,G$, can be written as:
\begin{equation}
\label{eqn:multiformula}
    \{\boldsymbol{y}_i \mid z_{ig}=1\}=(\boldsymbol{I}_K \otimes \boldsymbol{X})\boldsymbol{\alpha}_g +
    (\boldsymbol{I}_K \otimes \boldsymbol{W})\boldsymbol{\beta}_g + \boldsymbol{\epsilon}_i,
\end{equation}
where $\{z_{ig}\}_{g=1}^{G}$ are latent indicators as described in Section \ref{sec:multill}, $\boldsymbol{\alpha}_g=(\boldsymbol{\alpha}_{g1}^{\prime},\ldots,\boldsymbol{\alpha}_{gK}^{\prime})^{\prime}$ is a $2K$-vector of intercepts and slopes, $\boldsymbol{\beta}_g=(\boldsymbol{\beta}_{g1}^{\prime},\ldots,\boldsymbol{\beta}_{gK}^{\prime})^{\prime}$ is a $mK$-vector of basis function coefficients as described in Section \ref{sec:splineprior}, $\boldsymbol{I}_K$ is a $K \times K$ identity matrix and $\otimes$ denotes a tensor product. The matrix $\boldsymbol{X}$ is given by
$\boldsymbol{X}=\begin{pmatrix}
    1 & 1 & \ldots & 1 \\
    t_1 & t_2 & \ldots & t_n
\end{pmatrix}^{\prime}$ and the $m$ columns of the matrix $\boldsymbol{W}$ are smoothing splines basis functions as described in Section \ref{sec:splineprior}. We assume the error vector $\boldsymbol{\epsilon}_i$ follows a $\mbox{MVN}(\boldsymbol{0},\boldsymbol{\Psi}_g \otimes \boldsymbol{U})$ distribution, where $\boldsymbol{U}=\boldsymbol{I}_n$ is the $n \times n$ identity matrix, and $\boldsymbol{\Psi}_g=\mbox{diag}(\boldsymbol{\sigma}_g^2)$ is a $K \times K$ diagonal matrix with the error variances $\boldsymbol{\sigma}_g^2=(\sigma_{g1}^2,\ldots,\sigma_{gK}^2)^{\prime}$. We assume each subject has a common grid of time points across all $K$ entries, such that $\boldsymbol{X}$ and $\boldsymbol{W}$ are common to all subjects, although our proposed method can be generalized to the case where subjects are observed at different grids of time points. In addition, we assume $\boldsymbol{E}(\boldsymbol{y}_{ik}, \boldsymbol{y}_{ih})=\boldsymbol{0}_{n\times n}$ for $k\neq h$.

To simplify notation, we let $\boldsymbol{S}=[\boldsymbol{X} \ \boldsymbol{W}]$ and $\boldsymbol{\theta}_g=(\boldsymbol{\alpha}_{g1}^{\prime}, \boldsymbol{\beta}_{g1}^{\prime},\ldots,\boldsymbol{\alpha}_{gK}^{\prime}, \boldsymbol{\beta}_{gK}^{\prime})^{\prime}$. Equation~\eqref{eqn:multiformula} can then be rewritten as:
\begin{equation}
\label{eqn:multiformula2}
    \{\boldsymbol{y}_i\mid z_{ig}=1\}=(\boldsymbol{I}_K \otimes \boldsymbol{S})\boldsymbol{\theta}_g +
    \boldsymbol{\epsilon}_i.
\end{equation}

\subsection{Model for the mixing weights}
\label{sec:mixingweightsP1}

The mixture-of-experts model \citep{jacobs1991adaptive} is applied to form a covariate-guided structure for our proposed model, where the mixing weights are multinomial logits that are functions of selected covariates. As in \citet{sun2007multivariate}, the mixing weights are expressed as 
\begin{equation}
\label{eqn:unimixweight}
\pi_{ig}(\boldsymbol{V}_i)=\frac{\exp(\boldsymbol{V}_i^{\prime}\boldsymbol{\delta}_g+\zeta_{ig})}{\sum_{h=1}^G \exp(\boldsymbol{V}_i^{\prime}\boldsymbol{\delta}_h+\zeta_{ih})},
\end{equation}
where $\boldsymbol{V}_i=(1,V_{i1},\cdots,V_{iP})^{\prime}$ is a vector of length $(P+1)$ containing values of $P$ covariates for subject $i$, and  $\boldsymbol{\delta}_g=(\delta_{g0},\delta_{g1},\cdots,\delta_{gP})^{\prime}$ is the corresponding coefficient vector. For identifiability, we set $\boldsymbol{\delta}_G=\boldsymbol{0}$.  Equation~\eqref{eqn:unimixweight} differs slightly from the weights in the traditional mixture of experts model in that it includes a random term $\zeta_{ig}$ for each subject.  This term accounts for unmeasured factors beyond the observed covariates, and enhances model performance and inference of the mixing weights.

\subsection{Augmented likelihood}
\label{sec:multill}

To account for heterogeneity across subjects, we assume that the $k$th entry of the multivariate time series, $\boldsymbol{y}_{ik}$, comes from a mixture model with $G$ components, i.e.,
\begin{equation}
\label{eqn:mixtureform}
\boldsymbol{y}_{ik} \sim \sum_{g=1}^G \pi_{ig}f_{gk}(\boldsymbol{y}_{ik} \mid \boldsymbol{\mu}_{gk}, \sigma_{gk}^2\boldsymbol{I}_n),
\end{equation}
where $f_{gk}(\boldsymbol{y}_{ik} \mid \boldsymbol{\mu}_{gk}, \sigma_{gk}^2\boldsymbol{I}_n)$ is the probability density function of the multivariate normal distribution with mean vector $\boldsymbol{\mu}_{gk}=\boldsymbol{X} \boldsymbol{\alpha}_{gk} + \boldsymbol{W} \boldsymbol{\beta}_{gk}$ and covariance matrix $\sigma_{gk}^2 \boldsymbol{I}_n$ for the $g$th component and the $k$th entry. The $\pi_{ig}$ are mixing weights that depend on covariates as described in Section \ref{sec:mixingweightsP1}.

As is common in mixture models, augmenting the likelihood with latent variables indicating the component from which a time series originates simplifies the computation greatly \citep{dempster1977maximum}. In particular, let $z_{ig}=1$ if the $i$th multivariate time series belongs to the $g$th component and $z_{ig}=0$, otherwise. Let $\boldsymbol{y}=(\boldsymbol{y}_1,\ldots,\boldsymbol{y}_N)^{\prime}$ be  all observed multivariate time series and $\boldsymbol{\Theta}_{gk}$ be the aggregation of all parameters for component $g$ and entry $k$. The parameter vector for all components and all entries is then denoted by $\boldsymbol{\Theta}= (\boldsymbol{\Theta}_{11}^{\prime},\ldots,\boldsymbol{\Theta}_{GK}^{\prime})^{\prime}$.  The augmented likelihood of all $N$ multivariate time series is given by
\begin{equation}
\label{eqn:multifulllikelihood}
L(\boldsymbol{\Theta} \mid \boldsymbol{y},Z)=\prod_{i=1}^N\prod_{g=1}^G \Big[ \pi_{ig} \prod_{k=1}^K f_{gk}(\boldsymbol{y}_{ik}\mid\boldsymbol{\Theta}_{gk})\Big]^{z_{ig}},
\end{equation}
where $f_{gk}(\boldsymbol{y}_{ik} \mid \boldsymbol{\Theta}_{gk})$ is the probability density function as appeared in the \eqref{eqn:mixtureform}. From Bayes' rule, the distribution of the latent indicators $z_{ig}$ is given by
\begin{equation}
    \label{eqn:multilatentindi}
    p(z_{ig}=1\mid\boldsymbol{y}, \boldsymbol{S},\boldsymbol{\Theta},\pi_{ig})= \frac{\pi_{ig}\prod_{k=1}^K f_{gk}(\boldsymbol{y}_{ik}\mid\boldsymbol{\Theta}_{gk})}{\sum_{h=1}^G \pi_{ih}\prod_{k=1}^K f_{hk}(\boldsymbol{y}_{ik}\mid\boldsymbol{\Theta}_{hk})}.
\end{equation}

\section{Priors}
\label{sec:prior}

In this section, the priors on the model parameters are introduced.

\subsection{Smoothing splines prior}
\label{sec:splineprior}

The conditional expectation of a mixture component in model \eqref{eqn:mixtureform} is given by 
$E(\boldsymbol{y}_{ik} \mid z_{ig}=1)=\boldsymbol{X} \boldsymbol{\alpha}_{gk} + \boldsymbol{W} \boldsymbol{\beta}_{gk}$. 
We place a smoothing spline prior on $\boldsymbol{\beta}_{gk}$ and let $\boldsymbol{\mathcal{H}}_{gk}=\boldsymbol{W} \boldsymbol{\beta}_{gk}$, where $\boldsymbol{\mathcal{H}}_{gk}=\big[\mathcal{H}_{gk}(t_1),\ldots,\mathcal{H}_{gk}(t_n)\big]^{\prime}$ is a zero-mean Gaussian process with variance covariance matrix $\tau^2_{gk}\boldsymbol{\Phi}$ \citep{wahba1980automatic,wood2002bayesian}, such that $\text{cov}\big[\mathcal{H}_{gk}(t_{r}),\mathcal{H}_{gk}(t_h)\big]= \tau_{gk}^2\phi_{r h}$,  $\tau_{gk}^2$ is a smoothing parameter for component $g$ and entry $k$, and the $(r,h)$th element of $\boldsymbol{\Phi}$ is given by $\phi_{r h}=\frac{1}{2}t_{r}^2(t_h-\frac{t_{r}}{3})$ for $t_{r} \leq t_h$. The matrix $\boldsymbol{\Phi}$ is common to all subjects since all entries of the multivariate time series are observed at common time points.

As seen above, the matrix $\boldsymbol{\Phi}$ is $n \times n$, and to avoid the computational burden for large $n$, a low-rank approximation is often adopted. To facilitate this approximation, we obtain basis functions via the spectral decomposition of $\boldsymbol{\Phi}$, as has been proposed in \citet{wood2002bayesian} and used in \citet{rosen2009local, rosen2012adaptspec, krafty2011functional}. In particular, the matrix $\boldsymbol{W}$ consists of $m$ basis functions evaluated at times $t_1,\ldots,t_n$, and $\boldsymbol{\beta}_{gk}$ is an $m$-dimensional vector of basis function coefficients. These basis functions are obtained by applying the spectral decomposition to $\boldsymbol{\Phi}$ such that $\boldsymbol{\Phi}=\boldsymbol{Q}\boldsymbol{\Gamma}\boldsymbol{Q}^T$, where $\boldsymbol{Q}$ is the matrix of eigenvectors of $\boldsymbol{\Phi}$, and $\boldsymbol{\Gamma}$ is a diagonal matrix containing the eigenvalues of $\boldsymbol{\Phi}$. We then let the design matrix  $\boldsymbol{W}=\boldsymbol{Q}\boldsymbol{\Gamma}^{1/2}$ and place a normal prior $N(0,\tau^2_{gk}\boldsymbol{I}_n)$ on $\boldsymbol{\beta}_{gk}$, which leads to $\boldsymbol{\mathcal{H}}_{gk}$ or $\boldsymbol{W}\boldsymbol{\beta}_{gk} \sim N(\boldsymbol{0},\tau^2_{gk}\boldsymbol{\Phi})$ as mentioned above.

By using the low-rank approximation, the number of columns of $\boldsymbol{W}$ is reduced from $n$ to $m$ ($m<n$), which greatly reduces the computational burden without sacrificing the model fit \citep{wahba1980automatic, wood2006generalized}. \citet{eubank1999nonparametric} indicated that the eigenvalues in the diagonal matrix $\boldsymbol{\Gamma}$ decay rapidly as $m$ increases. Thus, we can achieve a good approximation by selecting a relatively small number $m$ of basis functions. The number of basis functions $m$ is set to $10$ in simulation studies as described in Section \ref{sec:sim}, which has been shown (\citealt{krafty2011functional}) to explain more than $98\%$ of the total variability.

The prior on $\boldsymbol{\theta}_g$ is thus $\boldsymbol{\theta}_g \sim N(\boldsymbol{0},\boldsymbol{D}_g)$, where $\boldsymbol{D}_g=$ diag$(\sigma_{\alpha 1}^2\boldsymbol{1}_2,\ \tau_{g1}^2\boldsymbol{1}_m, \ \ldots \ ,\sigma_{\alpha K}^2\boldsymbol{1}_2,\ \tau_{gK}^2\boldsymbol{1}_m)$ is the covariance matrix of $\boldsymbol{\theta}_g$. The vector $(\sigma_{\alpha 1}^2,\ldots,\sigma_{\alpha K}^2)^{\prime}$ contains fixed prior variances for the regression coefficients $\boldsymbol{\alpha}_{gk}$, common to all components and entries. In particular, we fix the common prior variance $\sigma_{\alpha}^2=100$. The vector $\boldsymbol{\tau}_g^2=(\tau_{g1}^2,\ldots,\tau_{gK}^2)^{\prime}$ contains the smoothing parameters for the $g$th mixture component and $\boldsymbol{1}_m$ is an $m$-vector of ones. We assume independence between the regression coefficients $\boldsymbol{\alpha}_{gk}$ and the basis function coefficients $\boldsymbol{\beta}_{gk}$.

\subsection{Priors on the smoothing parameters}
\label{sec:priorsmoothpam}

We assume the smoothing parameters $\boldsymbol{\tau}_g^2=(\tau_{g1}^2,\ldots,\tau_{gK}^2)^{\prime}$ vary across components $g$ and entries $k$. Although the most common choice for the prior on a variance parameter is the inverse gamma distribution, \citet{gelman2006prior} and \citet{wand2011mean} suggested that a half-$t$ prior on the standard deviation can reflect lack of information on a scale parameter. The half-$t$ is a family of heavy-tailed distributions and has a good shrinkage performance. It can be expressed as a scale mixture of inverse gamma random variables using a latent variable which follows an inverse gamma distribution \citep{wand2011mean}. Thus, we assume a half-$t$ distribution such that $\tau_{gk} \sim t_{\nu_{\tau}}^+(0,A_{\tau})$, where $\nu_{\tau}$ is a degrees of freedom parameter, and $A_{\tau}$ is a scale parameter. We set $\nu_{\tau}=3$ and $A_{\tau}=10$ for all components and entries.

\subsection{Priors on the error variances}
\label{sec:priorvariance}

We assume $\sigma_{gk} \stackrel{\text{i.i.d}}{\sim} t_{\nu_{\sigma}}^+(0,A_{\sigma})$ and set $\nu_{\sigma}=3$ and $A_{\sigma}=10$ for all components and entries.

\subsection{Priors on the logistic parameters and the variances of random intercepts}
\label{sec:priorlogistic}

This section provides details on the prior distributions placed on the parameters of the logistic weights \eqref{eqn:unimixweight}. For ease of notation, we denote $\boldsymbol{\delta}_g^*=(\boldsymbol{\delta}_g^T, \boldsymbol{\zeta}_g^T)^T$, where $\boldsymbol{\zeta}_g=(\zeta_{1g},\cdots,\zeta_{Ng})^T$, $g=1,\ldots,G$. We let $\boldsymbol{V}_i^*=(\boldsymbol{V}_i^{\prime},\boldsymbol{e}_i^{\prime})^{\prime}$ where $\boldsymbol{e}_i$ is a vector of all zeros except for a single $1$ in the $i$th position, and $\boldsymbol{V}^*$ is a matrix consisting of the rows $\boldsymbol{V}_i^{*T}$, $i=1,\ldots,N$.  Gaussian  priors are placed on the logistic parameters, i.e., $\boldsymbol{\delta}_g^* \sim N(\boldsymbol{0},\boldsymbol{B}_g)$, where $\boldsymbol{B}_g=\rm {diag} (\sigma_{\delta g}^2 \boldsymbol{1}_{P+1},\ \kappa^2_{\zeta g} \boldsymbol{1}_N)$, and the priors on the random intercepts satisfy $\boldsymbol{\zeta}_g \sim N(\boldsymbol{0}, \kappa^2_{\zeta g}\boldsymbol{I}_N)$. As for the hyperparameters, we assume $\sigma_{\delta g}^2=10$ for all components and covariates, and $\kappa_{\zeta g} \sim t_{\nu_{\kappa}}^+(0,A_{\kappa})$, where $\nu_{\kappa}=3$ and $A_{\kappa}=10$ for all components.

To sample the logistic parameters, \citet{polson2013bayesian} proposed a data augmentation scheme incorporating P\'{o}lya-Gamma latent variables, which facilitates Gibbs steps. Details on sampling the logistic parameters are provided in the Supplementary Material.

\section{Sampling scheme}
\label{sec:schemeP2}

This section outlines the Gibbs steps for sampling from the conditional posterior distributions of all the model parameters. More details are given in Supplementary Material.

\subsection{Gibbs sampling steps}
\label{sec:GibbsP2}

Letting $\ell$ denote the current Gibbs sampling iteration, parameter values at the $(\ell+1)$th iteration are drawn according to the following steps.

\begin{enumerate}
        \item Draw $\boldsymbol{\theta}_{gk}^{(\ell+1)}$ from $ (\boldsymbol{\theta}_{gk}^{(\ell+1)}\mid\boldsymbol{y},\boldsymbol{S}, \tau_{gk}^{2(\ell)}, \sigma_{gk}^{2 (\ell)}) \sim N(\boldsymbol{u}_{gk},\sigma_{gk}^2\boldsymbol{\Lambda}_{gk})$, where $\boldsymbol{u}_{gk}$ and $\boldsymbol{\Lambda}_{gk}$ are mean vectors and covariance matrices.

        \item \label{enum:sig} Draw $\sigma_{gk}^{2 (\ell+1)}$ from $(\sigma_{gk}^{2 (\ell+1)}\mid\boldsymbol{\epsilon}_{igk}^{(\ell+1)}, a_{\sigma_{gk}}^{(\ell+1)}) \sim IG \Big((n N_g^{(\ell)}+\nu_{\sigma})/2,\sum_{i=1}^Nz_{ig}\boldsymbol{\epsilon}_{igk}^{\prime}\boldsymbol{\epsilon}_{igk}/2+\nu_{\sigma}/a_{\sigma_{gk}}\Big)$, where $N_g^{(\ell)}$ is the current number of subjects in the $g$th component, $\boldsymbol{\epsilon}_{igk}$ is the error vector for the $g$th component, the $i$th subject and the $k$th entry, and $a_{\sigma_{gk}}$ is a latent variable in the $IG$ scale mixture underlying the half-$t$ distribution. 

        \item\label{enum:tau} Draw $\tau_{gk}^{2 (\ell+1)}$ from $(\tau_{gk}^{2 (\ell+1)}\mid\boldsymbol{\beta}_{gk}^{(\ell+1)}, a_{\tau_{gk}}^{(\ell+1)}) \sim IG \Big((\nu_{\tau}+m)/2,\boldsymbol{\beta}_{gk}^{\prime}\boldsymbol{\beta}_{gk}/2+\nu_{\tau}/a_{\tau_{gk}}\Big)$, where $a_{\tau_{gk}}$ is a latent variable as in \ref{enum:sig}. 

        \item Draw $\boldsymbol{\delta}_g^{*(\ell+1)}$ from $ (\boldsymbol{\delta}_g^{*(\ell+1)} \mid \boldsymbol{V^*}, z_{ig}^{(\ell)}, \omega_{ig}^{(\ell+1)}, \kappa^{2 (\ell)}_{\zeta g}) \sim N(\boldsymbol{M}_g,\boldsymbol{\Sigma}_g)$, where $\omega_{ig}^{(\ell+1)}$ is a P\'{o}lya-Gamma latent variable in the augmentation described in Section~\ref{sec:priorlogistic}.

        \item Draw $\kappa^{2 (\ell+1)}_{\zeta g}$ from $(\kappa^{2 (\ell+1)}_{\zeta g} \mid\boldsymbol{\zeta}_g^{(\ell+1)}, a_{\kappa_g}^{(\ell+1)}) \sim IG \Big(\nu_{\kappa}/2,\boldsymbol{\zeta}_g^{\prime}\boldsymbol{\zeta}_g/2+(\nu_{\kappa}+N)/a_{\kappa_g}\Big)$, where $a_{\kappa_g}$ is a latent variable as in \ref{enum:sig} and \ref{enum:tau}.

        \item The mixing weights $\pi_{ig}^{(\ell+1)}$ are obtained by computing $p(\pi_{ig}^{(\ell+1)} \mid \boldsymbol{V}^*,\boldsymbol{\delta}_g^{*(\ell+1)},z_{ig}^{(\ell)})$ from Equation \eqref{eqn:unimixweight}.

        \item Draw $z_{ig}^{(\ell+1)} \sim p(z_{ig}^{(\ell+1)}=1 \mid \boldsymbol{y}, \boldsymbol{S},\boldsymbol{\theta}_{gk}^{(\ell+1)},\sigma_{gk}^{2 (\ell+1)},\pi_{ig}^{(\ell+1)})$ according to Equation \eqref{eqn:multilatentindi}.
    \end{enumerate}

\subsection{Selecting the number of components}
\label{sec:selectcomp}

\citet{spiegelhalter2002bayesian} suggested the use of the deviance information criterion (DIC) for model selection based on the effective number of parameters. \citet{gelman2003bayesian} introduced an alternative measure of effective number of parameters based on the variance of the log predictive density across MCMC iterations. This measure is robust and more accurate than the original one. Moreover, it has the advantages of always being positive and invariant to reparameterizations \citep{gelman2003bayesian}.

In this paper, we use DIC to select the number of components for our proposed mixture model.

\section{Simulation studies}
\label{sec:sim}

To demonstrate the performance of the proposed method, we conduct simulation studies by generating data sets from the proposed model under two scenarios: two-component mixture ($G=2$) of trivariate time series ($K=3$) and four-component mixture ($G=4$) of bivariate time series ($K=2)$. We simulate $100$ replicates in each simulation setting with $N=150$ time series of length $n=50$.  A total of $20,000$ Gibbs sampling iterations are run with a burn-in of $4,000$. In all simulation settings, the hyperparameters are assigned the same values, given in Section \ref{sec:prior}.

\subsection{Two-component trivariate model}
\label{simtwocomp}

In this scenario, we consider the two-component trivariate model. From Equation~\eqref{eqn:multiformula},  the $g$th component of the proposed mixture model is given by
\begin{equation} 
\label{eqn:simformula}
\{ \boldsymbol{y}(t_j) \mid z_{ig}=1 \} = \boldsymbol{\alpha}_{0g}+ 
\boldsymbol{\alpha}_{1g}t_j+\sum_{q=1}^m w_q(t_j) \boldsymbol{\beta}_{gq}+\boldsymbol{\epsilon}_{gt_j}, \quad j=1,\ldots,n,\;\;g=1,\ldots,G,
\end{equation}
where $\boldsymbol{y}(t_j)$ is the trivariate time series evaluated at time $t_j$, $\boldsymbol{\alpha}_{01}= (1,-3,-2)^{\prime}$, $\boldsymbol{\alpha}_{02}=(5,4,3)^{\prime}$ and $\boldsymbol{\alpha}_{11}=(-2,2,0.5)^{\prime}$, $\boldsymbol{\alpha}_{12}=(1,-1,-0.5)^{\prime}$ are independent intercepts and slopes for each component, respectively. The vector $\boldsymbol{\beta}_{gq}$ consists of the $q$th spline coefficients of all variates for component $g$, and $w_q(t_j)$ is the $q$th spline basis function evaluated at time $t_j$. The $\boldsymbol{\epsilon}_{gt_j}$ are independent zero-mean error terms, distributed as $\boldsymbol{\epsilon}_{gt_j} \sim \text{MVN} \Big(\boldsymbol{0}, {\rm diag} (\sigma_{g1}^2,\sigma_{g2}^2, \sigma_{g3}^2)\Big)$, where $\sigma_{1}^2=(\sigma_{11}^2,\sigma_{12}^2,\sigma_{13}^2)^{\prime}=(3,5,4.5)^{\prime}$ and $\sigma_{2}^2=(\sigma_{21}^2,\sigma_{22}^2,\sigma_{23}^2)^{\prime}=(4,3.5,4)^{\prime}$. The smoothing parameters are set to $\tau_{1}^2=(\tau_{11}^2,\tau_{12}^2,\tau_{13}^2)^{\prime}=(3.5,5,8.5)^{\prime}$ and $\tau_{2}^2=(\tau_{21}^2,\tau_{22}^2,\tau_{23}^2)^{\prime}=(6,2.5,1.5)^{\prime}$.

We investigate the performance of the trajectory and logistic parameter (see Equation~\eqref{eqn:unimixweight}) estimates. For the former, we calculate the averaged root square error (ARSE) of each mixture component $g$
\begin{equation*}
    \text{ARSE}_g =\sqrt{\frac{1}{nK}\sum_{j=1}^n \sum_{k=1}^K \Big[ \mu_{gk}(t_j)-\hat{\mu}_{gk}(t_j) \Big]^2},
\end{equation*}
where $\mu_{gk}(t_j)$ is the expectation of $y_k(t_j)$ according to the $g$th component, and $y_k(t_j)$ is the $k$th entry of the time series evaluated at time $t_j$. The $\hat{\mu}_{gk}(t_j)$ are the estimated posterior means of $\mu_{gk}(t_j)$ for $k=1,\ldots,K$ and $j=1,\ldots,n$.

To handle a potential label switching across mixture components, we compute $\text{ARSE}_g$ as the minimum value across all components, by using the estimate of the $g$th component and the truth of each group, $g=1,\ldots,G$. After obtaining correct component labels by evaluating ARSE, we also report the averaged bias (A-bias) and the variance of the bias (V-bias) of each mixture component $g$, where
\begin{equation*}
    \text{A-bias}_g=\frac{1}{nK}\sum_{j=1}^n \sum_{k=1}^K \Big[ \hat{\mu}_{gk}(t_j)-\mu_{gk}(t_j) \Big],
\end{equation*}
and $\text{V-bias}_g$ is computed by calculating the sample variance of the bias over entries and time points.

\begin{figure}[!p]
\centering\includegraphics[width=1.0\textwidth]{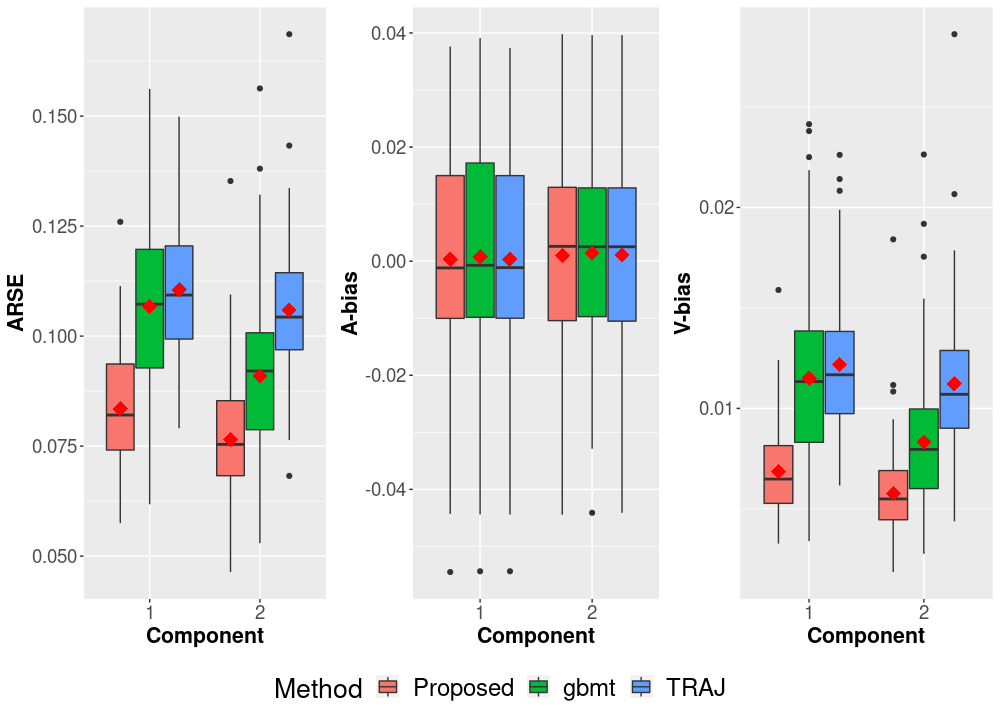}
\caption{Boxplots of the averaged root square error (ARSE), the averaged bias (A-bias) and the variance of bias (V-bias) of estimated trajectories for each component from $100$ replicates of $150$ two-component trivariate time series of length $50$. The proposed method was compared to \texttt{R} package \texttt{gbmt} and \texttt{TRAJ} procedure in \texttt{SAS}. The diamond markers denote the mean statistics of each method and component.}
\label{fig:sim2g}
\end{figure}

For each replicate, time series trajectories are estimated by three methods: the proposed method, the $\texttt{R}$ package $\texttt{gbmt}$ \citep{magrini2022assessment} and the $\texttt{TRAJ}$ procedure in \texttt{SAS} \citep{nagin2018group}. Boxplots of ARSE, A-bias and V-bias of each component are given in Figure \ref{fig:sim2g}. Notably, $\texttt{TRAJ}$ is able to fit a regression spline model by treating basis functions as time-varying covariates, while $\texttt{gbmt}$ is only able to fit a cubic model. Our proposed method fits a penalized spline model under the Bayesian framework and is able to outperform both $\texttt{gbmt}$ and $\texttt{TRAJ}$ in terms of ARSE and V-bias for both components. A-biases are close to zero and comparable for all three methods. These findings demonstrate that all three methods are able to achieve a reasonable fit to group-based trajectories since bias over the entire time series is close to zero. Our proposed method is able to obtain more precise estimates of trajectories as is evident from the smaller V-biases.  

To evaluate the performances of the logistic parameters, we compute the root mean squared error (RMSE) for each logistic parameter using the proposed method and $\texttt{TRAJ}$. Notably, $\texttt{gbmt}$ is not able to incorporate covariates into the computation of mixing weights. Results of RMSEs of each logistic parameter are given in Table \ref{Tablelog2g}. We also compare RMSEs between the proposed method and $\texttt{TRAJ}$ under four settings of different combinations of $N=150, 250$ and $n=50, 70$. Our proposed method yields smaller RMSEs of the logistic parameters in all cases, especially for the intercept $\delta_0$ and the first covariate $\delta_1$. This is to be expected since $\texttt{TRAJ}$ uses a multinomial logistic model, which may result in inflated parameter estimates in cases of unbalanced outcomes or perfect separation, while our proposed method is able to obtain a shrinkage result using the penalization method.

\subsection{Four-component bivariate model}
\label{simtwocomp}

In this scenario, we consider the four-component bivariate model whose $g$th component is given in Equation \eqref{eqn:simformula}, where the values of the intercepts and slopes are $\boldsymbol{\alpha}_{01}= (1,-2)^{\prime}$, $\boldsymbol{\alpha}_{02}= (5,3)^{\prime}$, $\boldsymbol{\alpha}_{03}= (-3,5.5)^{\prime}$, $\boldsymbol{\alpha}_{04}= (4,-1)^{\prime}$,   $\boldsymbol{\alpha}_{11}= (-3,0)^{\prime}$, $\boldsymbol{\alpha}_{12}= (2,-3.5)^{\prime}$, $\boldsymbol{\alpha}_{13}= (2.5,2)^{\prime}$ and $\boldsymbol{\alpha}_{14}= (-3,1.5)^{\prime}$. By analogy to the two-component trivariate model, the errors $\boldsymbol{\epsilon}_{gt_j}$ are independent zero-mean bivariate Gaussian random variables, distributed as $\boldsymbol{\epsilon}_{gt_j} \sim \text{MVN} \Big(\boldsymbol{0}, {\rm diag} (\sigma_{g1}^2,\sigma_{g2}^2)\Big)$, where $\sigma_{1}^2=(\sigma_{11}^2,\sigma_{12}^2)^{\prime}=(6,9)^{\prime}$, $\sigma_{2}^2=(\sigma_{21}^2,\sigma_{22}^2)^{\prime}=(8,7.5)^{\prime}$, $\sigma_{3}^2=(\sigma_{31}^2,\sigma_{32}^2)^{\prime}=(10,6.5)^{\prime}$ and $\sigma_{4}^2=(\sigma_{41}^2,\sigma_{42}^2)^{\prime}=(7,8.5)^{\prime}$.

\begin{figure}[!p]
\centering\includegraphics[width=1.0\textwidth]{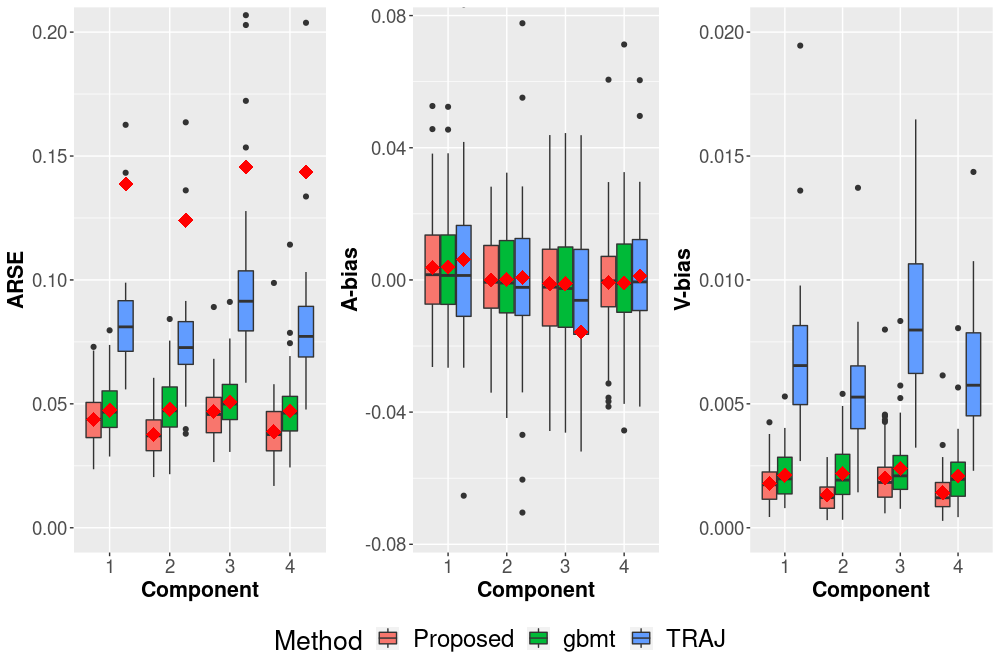}
\caption{Boxplots of the averaged root square error (ARSE), the averaged bias (A-bias) and the variance of bias (V-bias) of estimated trajectories for each component from $100$ replicates of $150$ four-component bivariate time series of length $50$. The proposed method was compared to \texttt{R} package \texttt{gbmt} and \texttt{TRAJ} procedure in \texttt{SAS}. The diamond markers denote the mean statistics of each method and component. All boxplots are zoomed in for better visualization. }
\label{fig:sim4g}
\end{figure}

The performances of the estimated trajectories and logistic parameters for this scenario are displayed in Figure \ref{fig:sim4g} and Table \ref{Tablelog4g1}. As in the first scenario, our proposed method outperforms both \texttt{gbmt} and \texttt{TRAJ} in terms of ARSE and V-bias for all components. Notably, \texttt{TRAJ} fails to yield precise estimates in several replicates and thus results in larger mean ARSE and V-bias. In terms of the logistic parameters, the proposed method performs well with smaller RMSEs in almost all cases, especially for $\delta_0$ and $\delta_1$. More simulation results based on different values of $N$ and $n$ under the two scenarios considered above are presented in the Supplementary Material.

\section{Real data application}
\label{sec:realdataapp}

We apply our proposed method to the analysis of the fNIRS still-face study introduced in Section \ref{sec:motivating}. Six covariates are considered in our covariate-guided model, including Infant Behavior Questionnaire-Revised negative emotionality (IBQ-NE) score, Infant Behavior Questionnaire-Revised effortful control (IBQ-EC) score, gestational age (in Days), infant age (in Months), head circumference (in cm) and sex. All continuous covariates are centered and scaled. We set the number of basis functions at $m=20$ and run a total of $30,000$ Gibbs iterations with a burn-in period of $6,000$. The values of the hyperparameters are the same as the ones used in the simulation studies.

The IBQ-NE construct combines data from the following subscales: Sadness, Distress to Limitations, Fear, and Falling Reactivity/Rate of Recovery from Distress. IBQ-EC refers to the ability to inhibit a dominant response to perform a subdominant one and has been shown to be protective against a myriad of difficulties \citep{gartstein2013origins}. Finally, the data consist of 79 subjects with complete fNIRS and covariate values. We present results based on analyzing one set of four-channels. Additional results based on analyzing another set of four channels and all channels are given in the Supplementary Material. The four channels are S1D1, S2D2, S5D3 and S6D4. Channels S1D1 and S5D3 are in the central prefrontal region, while channels S2D2 and S6D4 are in the left and right prefrontal region, respectively. We fit our proposed model with the number of components varying from 2 to 6. Based on values of DIC introduced in Section \ref{sec:selectcomp}, the two-component model is selected as the best model for this four-channel analysis.

\begin{figure}[!p]		
\centering\includegraphics[width=1.0\textwidth]{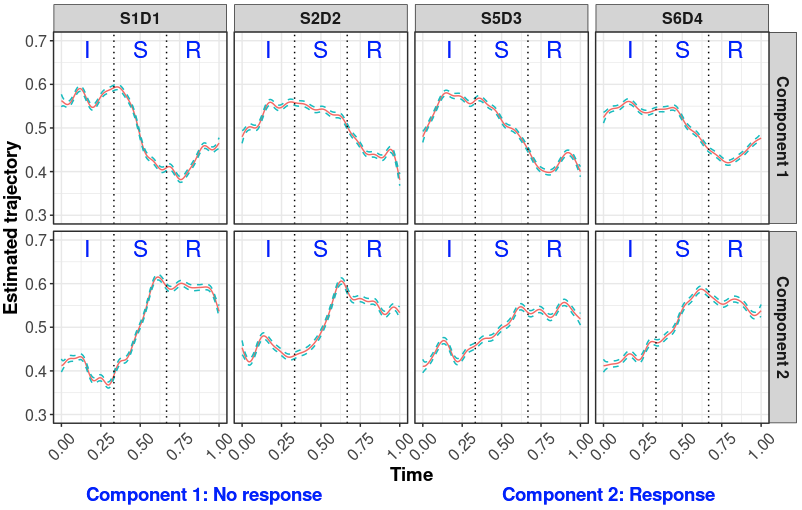}
\caption{Estimated trajectories of the two-component model with four selected channels. \textbf{I}: Interact \textbf{S}: Still-face \textbf{R}: Recovery. Red curves are posterior mean and two green dashed curves are 95\% pointwise credible intervals.}
\label{fig:fNIRS4g1}
\end{figure}

\begin{figure}[!p]		
\centering\includegraphics[width=0.8\textwidth]{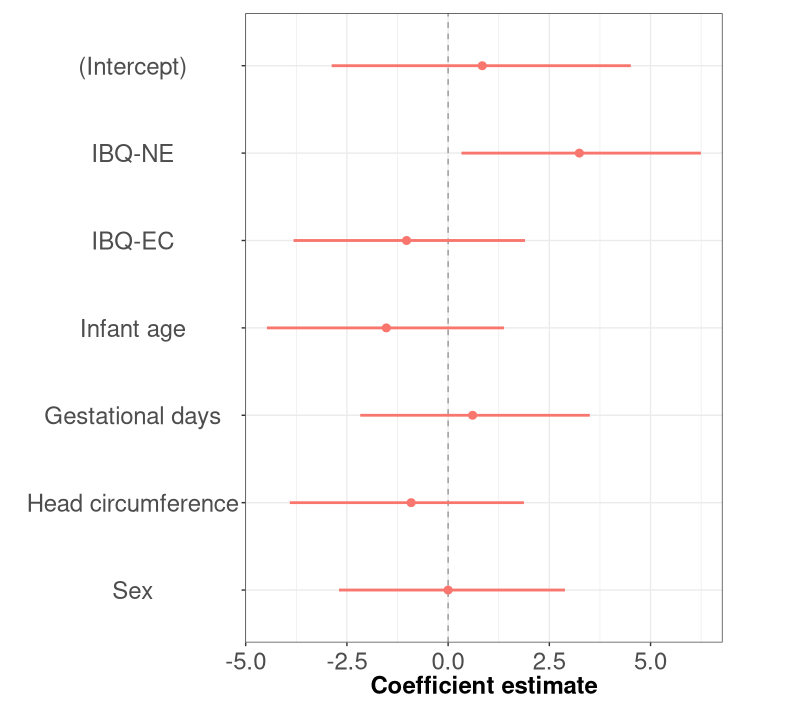}
\caption{Logistic coefficient estimates and 95\% credible intervals for each covariate of the two-component model.}
\label{fig:fNIRS4g2}
\end{figure}

Figure \ref{fig:fNIRS4g1} presents the estimated trajectories of the two-component model fitted to the four channels. We are interested in brain activation signals in the still-face period while the interact period is used as the reference level. For component 1, a decreasing trajectory is observed for the still-face period in all four channels. In contrast, an increasing trend is observed for the still-face period in all four channels for component 2. After fitting the mixture model and finding above trajectory patterns, we define component 1 as the no response component and component 2 as the response component based on trajectory patterns in the still-face period. Figure \ref{fig:fNIRS4g2} displays the logistic parameter estimates for all covariates in the 2-component model, where component 2 is used as the reference. There is evidence that IBQ-NE scores differ between the two components as its 95\% credible interval does not include zero. A positive coefficient of IBQ-NE indicates that a higher IBQ-NE score is associated with component 1, which has decreased brain activation levels in the still-face period for all four channels. Though other logistic coefficients have 95\% credible intervals that include zero, the negative posterior mean estimate of the IBQ-EC score could still indicate that a high IBQ-EC is associated with an increased brain activation as shown for component 2. These conclusions are consistent with findings in \citet{gartstein2013origins} that IBQ-NE is negatively associated with IBQ-EC. \citet{enlow2016infant} reported a negative association between activity level and IBQ-NE among infants whose families encourage a high level of activities. Furthermore, a negative posterior mean of logistic coefficient of infant age suggests that younger infant tends to have a decreasing brain activation level in the still-face period.

\section{Discussion}
\label{sec:discussion}

The proposed covariate-guided Bayesian mixture of spline experts model aims to perform a model-based clustering of multivariate time series from multiple subjects. The mixture components in this model are penalized splines, and the mixing weights incorporate covariates. Our proposed method is compared to two commonly used methods through simulation studies which demonstrate a better performance of our method under different scenarios. We apply our proposed method to a fNIRS still-face study and find distinct patterns of components of time series trajectories, as well as an association between IBQ-NE score and a pattern of decreased brain activity in the still-face period. To the best of our knowledge, this is the first still-face study using fNIRS whose purpose is to identify trajectory components.

Our proposed method has some limitations. First, as in any mixture models, label switching may occur, especially in the real-data application. We have adopted the Equivalence Classes Representatives (ECR) algorithm proposed by \citet{papastamoulis2010artificial} to make the components interpretable, but other methods may be considered. Second, the proposed method assumes independence among the entries of the time series and does not allow spatial dependence. Spatial correlations of fNIRS are correlations among fNIRS channels based on the placements and locations of each source and detector. An extension to a multilevel multivariate model would be possible by considering spatial correlations among time series entries. Lastly, our proposed method uses DIC to select the number of components which might be sub-optimal. Bayesian model averaging and reversible jump MCMC (RJMCMC) methods could be considered, but trans-dimensional sampling methods would pose challenges in providing interpretable components.

\section{Software}
\label{sec:software}

Software in the form of R codes, together with an example data, is available at \url{https://github.com/HaoyiFu1993/CBMOSE}.

\bibliographystyle{biorefs}
\bibliography{reference}

\begin{table}[!p]
\centering
\caption{Root mean square errors (RMSEs) of each logistic parameter for the two-component trivariate model from $100$ replicates of $N$ two-component trivariate time series of length $n$. RMSEs of the proposed method were compared to \texttt{TRAJ} procedure in \texttt{SAS}. Parameters $\delta_0$, $\delta_1$, $\delta_2$ and $\delta_3$ are intercept, first, second and third logistic parameters, respectively. The true values of logistic parameters are $5,-3.5,1,0.1$, respectively}
\label{Tablelog2g}
{\tabcolsep=12pt
\begin{tabular}{@{}ccccccc@{}}
\hline
\multicolumn{1}{l}{n} & N & Method & $\delta_0$ & $\delta_1$ & $\delta_2$ & $\delta_3$ \\ \hline
\multirow{2}{*}{50} & \multirow{2}{*}{150} & Proposed & 0.89    & 0.52    & 0.29        & 0.32    \\ 
                    &                      & TRAJ     & 1.57    & 0.87    & 0.36        & 0.34    \\ \hline
\multirow{2}{*}{70} & \multirow{2}{*}{150} & Proposed & 0.86    & 0.50    & 0.29        & 0.31    \\  
                    &                      & TRAJ     & 1.55    & 0.86    & 0.36        & 0.34    \\ \hline
\multirow{2}{*}{50} & \multirow{2}{*}{250} & Proposed & 0.77    & 0.40    & 0.22        & 0.23    \\  
                    &                      & TRAJ     & 0.96    & 0.50    & 0.23        & 0.24    \\ \hline
\multirow{2}{*}{70} & \multirow{2}{*}{250} & Proposed & 0.77    & 0.41    & 0.22        & 0.23    \\ 
                    &                      & TRAJ     & 0.97    & 0.51    & 0.24        & 0.24    \\ \hline
\end{tabular}
}
\end{table}

\begin{table}[!p]
\centering
\caption{Root mean square errors (RMSEs) of each logistic parameter for the four-component bivariate model from $100$ replicates of $150$ four-component bivariate time series of length $50$. RMSEs of the proposed method were compared to \texttt{TRAJ} procedure in \texttt{SAS}. Parameters $\delta_0$, $\delta_1$, $\delta_2$ and $\delta_3$ are intercept, first, second and third logistic parameters, respectively. The fourth component was used as the reference component. The true values of logistic parameters are $5,-3.5,1,0.1$ (first component), $-4,2.5,-2,-0.2$ (second component), $3,-2,0.8,0.2$ (third component). C1, C2, C3 and C4 denote first, second, third and fourth component, respectively.}
\label{Tablelog4g1}
{\tabcolsep=12pt
\begin{tabular}{@{}clcccccc@{}}
\hline
\multicolumn{1}{l}{n} & N & Method & Comparison & $\delta_0$ & $\delta_1$ & $\delta_2$ & $\delta_3$ \\ \hline

\multirow{6}{*}{50} & \multirow{6}{*}{150} & \multirow{3}{*}{Proposed}      & C1 vs C4   & 0.81  & 0.53 & 0.30 & 0.39 \\ 
                             & & & C2 vs C4   & 1.11 & 0.46 & 0.42 & 0.36 \\  
                             & & & C3 vs C4   & 0.89 & 0.42 & 0.28 & 0.34 \\ \cline{3-8}
& & \multirow{3}{*}{TRAJ}   & C1 vs C4   & 1.20 & 0.74 & 0.35 & 0.41 \\  
                             & & & C2 vs C4   & 3.81 & 2.27 & 1.33 & 0.49 \\ 
                             & & & C3 vs C4   & 2.07 & 1.33 & 0.76 & 0.32 \\ \hline
\end{tabular}
}
\end{table}

\section{Supplemental material}
\subsection*{Appendix A: Details of the sampling scheme}
\label{sec:detailss}

As described in Section 5 of the paper, Gibbs sampling is used to facilitate Bayesian inference.   We denote by $\boldsymbol{\Theta}_{gk} = (\boldsymbol{\theta}_{gk}^{\prime}, \tau_{gk}^{2}, \sigma_{gk}^{2}, \boldsymbol{\delta}_g^{*\prime},\kappa^2_{\zeta g})^{\prime}$ the parameters for the $g$th component and the $k$th entry, and the parameters in this vector are drawn from the corresponding conditional posterior distributions. Let $\ell$ be the current Gibbs sampling iteration; detailed Gibbs sampling steps for drawing the parameters at the $(\ell+1)$th iteration are given below.

\begin{enumerate}
    
\item \textbf{Sampling the basis function coefficients}
    
For each component $g$ and time series entry $k$, based on the augmented likelihood in Section 3.3 and the priors on $\boldsymbol{\theta}_{gk}=(\boldsymbol{\alpha}_{gk}^{\prime},\boldsymbol{\beta}_{gk}^{\prime})^{\prime}$ described in Section 4.1, the conditional posterior distribution of $(\boldsymbol{\theta}_{gk}^{(\ell+1)} \mid\boldsymbol{y},\boldsymbol{S},\tau_{gk}^{2 (\ell)},\sigma_{gk}^{2 (\ell)})$ is:
    
\begin{equation*}
\begin{aligned}
p(\boldsymbol{\theta}_{gk}^{(\ell+1)} \mid\boldsymbol{y},\boldsymbol{S},& \tau_{gk}^{2 (\ell)},\sigma_{gk}^{2 (\ell)}) \propto p(\boldsymbol{y}\mid \boldsymbol{S},\boldsymbol{\theta}_{gk}^{(\ell+1)},\sigma_{gk}^{2 (\ell)}) \cdot p(\boldsymbol{\theta}_{gk}^{(\ell+1)} \mid\tau_{gk}^{2 (\ell)})  \\ &\propto \prod_{i=1}^N \Big\{(\sigma_{gk}^2)^{-n/2}\exp \big[-\frac{1}{2\sigma_{gk}^2}(\boldsymbol{y}_{ik} - \boldsymbol{S}\boldsymbol{\theta}_{gk})^{\prime}(\boldsymbol{y}_{ik} - \boldsymbol{S}\boldsymbol{\theta}_{gk})\big]\Big\}^{z_{ig}} \\
& \times \vert\boldsymbol{D}_{gk}\vert^{-1/2}\exp \Big(-\frac{1}{2}\boldsymbol{\theta}_{gk} \boldsymbol{D}_{gk}^{-1}\boldsymbol{\theta}_{gk} \Big) \\
& \propto \exp\Big\{-\frac{1}{2\sigma_{gk}^2} \big[\sum_{i=1}^N z_{ig}(\boldsymbol{y}_{ik} - \boldsymbol{S}\boldsymbol{\theta}_{gk})^{\prime}(\boldsymbol{y}_{ik} - \boldsymbol{S}\boldsymbol{\theta}_{gk}) + \boldsymbol{\theta}_{gk}^{\prime} \sigma_{gk}^2\boldsymbol{D}_{gk}^{-1}\boldsymbol{\theta}_{gk} \big] \Big\}\\
& \propto \exp\Big[ -\frac{1}{2\sigma_{gk}^2}(\boldsymbol{\theta}_{gk}-\boldsymbol{u}_{gk})^{\prime} (\boldsymbol{\Lambda}_{gk})^{-1}(\boldsymbol{\theta}_{gk}-\boldsymbol{u}_{gk}) \Big] \\
& \sim N(\boldsymbol{u}_{gk}, \sigma_{gk}^2\boldsymbol{\Lambda}_{gk}),
\end{aligned}
\end{equation*}
where $\boldsymbol{\Lambda}_{gk}=(N_g^{(\ell)} \boldsymbol{S}^{\prime}\boldsymbol{S}+\sigma_{gk}^2\boldsymbol{D}_{gk}^{-1})^{-1}$, $\boldsymbol{u}_{gk}=\boldsymbol{\Lambda}_{gk}\sum_{i=1}^N z_{ig}\boldsymbol{S}^{\prime}\boldsymbol{y}_{ik}$, $N_g^{(\ell)}$ is the current number of subjects in the $g$th component, $\boldsymbol{D}_{gk}= {\rm diag}(\sigma_{\alpha}^2\boldsymbol{1}_2,\tau_{gk}^2\boldsymbol{1}_m)$ is the prior covariance matrix for $\boldsymbol{\theta}_{gk}$. Hence, for each component $g$ and entry $k$, we draw $\boldsymbol{\theta}_{gk}^{(\ell+1)}$ from $(\boldsymbol{\theta}_{gk}^{(\ell+1)} \mid\boldsymbol{y},\boldsymbol{S}, \tau_{gk}^{2 (\ell)},\sigma_{gk}^{2 (\ell)}) \sim N(\boldsymbol{u}_{gk},\sigma_{gk}^2 \boldsymbol{\Lambda}_{gk})$.

\item \textbf{Sampling the error variances}
    
\citet{gelman2006prior} proposed using the half-$t$ distribution as the prior on scale parameters. We follow \citet{wand2011mean} and express the half-$t$ prior of Section 4.3 as a scale mixture of inverse Gamma distributions as follows  
\begin{equation*}
    (\sigma_{gk}^2\mid a_{\sigma_{gk}}) \sim IG \Big(\frac{\nu_{\sigma}}{2},\frac{\nu_{\sigma}}{a_{\sigma_{gk}}}\Big), a_{\sigma_{gk}} \sim IG\Big(\frac{1}{2},\frac{1}{A_{\sigma}^2}\Big).
\end{equation*}
Therefore, the conditional posterior distribution of the latent variable $a_{\sigma_{gk}}$ is
\begin{equation*}
 p(a_{\sigma_{gk}}^{(\ell+1)} \mid \sigma_{gk}^{2 (\ell)}) \propto \exp\Big[-\frac{1}{a_{\sigma_{gk}}} \Big(\frac{\nu_{\sigma}}{\sigma_{gk}^2}+\frac{1}{A_{\sigma}^2}\Big)\Big] \times (a_{\sigma_{gk}})^{-(\frac{1}{2}+1+\frac{\nu_{\sigma}}{2})},
\end{equation*}
which is $IG \Big(\frac{\nu_{\sigma}+1}{2},\frac{\nu_{\sigma}}{\sigma_{gk}^2}+\frac{1}{A_{\sigma}^2}\Big)$. Denoting by $\boldsymbol{\epsilon}_{igk}$ the error vector of time series $\boldsymbol{y}_{ik}$ for component $g$,  we have $\boldsymbol{\epsilon}_{igk}=\boldsymbol{y}_{ik}-\boldsymbol{S}\boldsymbol{\theta}_{gk}$, where $\boldsymbol{\epsilon}_{igk} \sim N(\boldsymbol{0},\sigma_{gk}^2\boldsymbol{I}_n)$. The conditional distribution of the error variance is
\begin{equation*}
\begin{aligned}
p(\sigma_{gk}^{2 (\ell+1)} & \mid \boldsymbol{\epsilon}_{igk}^{(\ell+1)},a_{\sigma_{gk}}^{(\ell+1)}) \propto p(\boldsymbol{\epsilon}_{igk}^{(\ell+1)} \mid \sigma_{gk}^{2 (\ell+1)}) \cdot p(a_{\sigma_{gk}}^{(\ell+1)} \mid \sigma_{gk}^{2 (\ell+1)}) \cdot p(\sigma_{gk}^{2 (\ell+1)}) \\
& \propto \prod_{i=1}^N \Big[(\sigma_{gk}^2)^{-\frac{n}{2}} \exp\Big(-\frac{1}{2\sigma_{gk}^2} \boldsymbol{\epsilon}_{igk}^{\prime} \boldsymbol{\epsilon}_{igk}\Big)\Big]^{z_{ig}} \times (\sigma_{gk}^2)^{-(\frac{\nu_{\sigma}}{2}+1)} \exp\Big(-\frac{\nu_{\sigma}}{\sigma_{gk}^2a_{\sigma_{gk}}}\Big) \\
& \propto (\sigma_{gk}^2)^{-(\frac{n}{2}N_g^{(\ell)}+\frac{\nu_{\sigma}}{2}+1)} \exp\Big[-\frac{1}{\sigma_{gk}^2} \Big(\frac{\sum_{i=1}^N z_{ig}\boldsymbol{\epsilon}_{igk}^{\prime} \boldsymbol{\epsilon}_{igk}}{2} + \frac{\nu_{\sigma}}{a_{\sigma_{gk}}}\Big)\Big],
\end{aligned}
\end{equation*}
which is $IG\Big(\frac{nN_g^{(\ell)}+\nu_{\sigma}}{2}, \frac{\sum_{i=1}^N z_{ig}\boldsymbol{\epsilon}_{igk}^{\prime} \boldsymbol{\epsilon}_{igk}}{2} + \frac{\nu_{\sigma}}{a_{\sigma_{gk}}}\Big)$. The sampling scheme proceeds by first sampling $(a_{\sigma_{gk}}^{(\ell+1)} \mid \sigma_{gk}^{2 (\ell)})$  and then $(\sigma_{gk}^{2 (\ell+1)} \mid \boldsymbol{\epsilon}_{igk}^{(\ell+1)},a_{\sigma_{gk}}^{(\ell+1)})$.

\item \textbf{Sampling the smoothing parameters}
    
The smoothing parameters $\tau_{gk}^2$ are drawn by analogy to the error variances. We first draw $(a_{\tau_{gk}}^{(\ell+1)} \mid \tau_{gk}^{2 (\ell)}) \sim IG \Big(\frac{\nu_{\tau}+1}{2},\frac{\nu_{\tau}}{\tau_{gk}^2}+\frac{1}{A_{\tau}^2} \Big)$. The conditional posterior distribution of the smoothing parameters is
\begin{equation*}
\begin{aligned}
p(\tau_{gk}^{2 (\ell+1)} \mid \boldsymbol{\beta}_{gk}^{(\ell+1)},a_{\tau_{gk}}^{(\ell+1)}) & \propto p(\boldsymbol{\beta}_{gk}^{(\ell+1)} \mid \tau_{gk}^{2 (\ell+1)}) \cdot p(a_{\tau_{gk}}^{(\ell+1)} \mid \tau_{gk}^{2 (\ell+1)}) \cdot p(\tau_{gk}^{2 (\ell+1)})  \\
& \propto (\tau_{gk}^2)^{-\frac{m+\nu_{\tau}}{2}} \exp{\Big[-\frac{1}{\tau_{gk}^2} \Big(\frac{\nu_{\tau}}{a_{\tau_{gk}}}+\frac{\boldsymbol{\beta}_{gk}^{\prime}\boldsymbol{\beta}_{gk}}{2}\Big)\Big]},
\end{aligned}
\end{equation*}
which is $IG\Big (\frac{\nu_{\tau}+m}{2}, \frac{\boldsymbol{\beta}_{gk}^{\prime}\boldsymbol{\beta}_{gk}}{2} + \frac{\nu_{\tau}}{a_{\tau_{gk}}}\Big)$. The sampling scheme proceeds by first sampling $(a_{\tau_{gk}}^{(\ell+1)} \mid \tau_{gk}^{2 (\ell)})$ and then $(\tau_{gk}^{2 (\ell+1)} \mid \boldsymbol{\beta}_{gk}^{(\ell+1)},a_{\tau_{gk}}^{(\ell+1)})$.

\item \textbf{Sampling the logistic parameters}
        
Let $\boldsymbol{\delta}_g^*=(\boldsymbol{\delta}_g^T, \boldsymbol{\zeta}_g^T)^T$ be the aggregation of the logistic parameters and all random intercepts for the $g$th component. Based on the logits of Section 3.2 and the corresponding priors described in Section 4.4, the conditional posterior distribution of $(\boldsymbol{\delta}_{g}^{* (\ell+1)} \mid \boldsymbol{V}^*, z_{ig}^{(\ell)},\kappa^{2 (\ell)}_{\zeta g})$ is 
\begin{equation*}
\begin{aligned}
p(\boldsymbol{\delta}_{g}^{* (\ell+1)}\mid \boldsymbol{V}^*, z_{ig}^{(\ell)},\kappa^{2 (\ell)}_{\zeta g}) & \propto p(z_{ig}^{(\ell)}=1 \mid \boldsymbol{V}^*, \boldsymbol{\delta}_{g}^{* (\ell+1)}) \cdot p(\boldsymbol{\delta}_{g}^{* (\ell+1)} \mid \kappa^{2 (\ell)}_{\zeta g}) \\
& =\prod_{i=1}^N \Big[\frac{\exp(\boldsymbol{V}_i^{*\prime}\boldsymbol{\delta}_{g}^*)}{\sum_{h=1}^G\exp(\boldsymbol{V}_i^{*\prime}\boldsymbol{\delta}_{h}^{*})}\Big]^{z_{ig}} p(\boldsymbol{\delta}_{g}^{* (\ell+1)} \mid \kappa^{2 (\ell)}_{\zeta g}),
\end{aligned}
\end{equation*}
where $\boldsymbol{V}^*=(\boldsymbol{V}_1^*,\ldots,\boldsymbol{V}_N^*)^{\prime}$ is a $N \times (P+1)$ matrix with $\boldsymbol{V}_i^*$ representing all covariates (including intercepts) for subject $i$. To sample from the posterior distribution of $p(\boldsymbol{\delta}_{g}^{* (\ell+1)}\mid \boldsymbol{V}^*, z_{ig}^{(\ell)},\kappa^{2 (\ell)}_{\zeta g})$, we adopt the P\'{o}yla-Gamma data augmentation strategy of \citet{polson2013bayesian} by introducing a latent variable $\omega_{ig}$ coming from the P\'{o}lya-Gamma distribution. Thus, the conditional posterior distributions of the logistic parameters are
\begin{equation*}
\begin{aligned}
p(\boldsymbol{\delta}_{g}^{* (\ell+1)} \mid \boldsymbol{V}^*, z_{ig}^{ (\ell)},\omega_{ig}^{(\ell+1)},\kappa^{2 (\ell)}_{\zeta g}) &\propto p(z_{ig}^{ (\ell)} = 1 \mid \boldsymbol{V}^*, \omega_{ig}^{(\ell+1)}, \boldsymbol{\delta}_{g}^{* (\ell+1)}) \cdot p(\omega_{ig}^{(\ell+1)} \mid \boldsymbol{V}^*, \boldsymbol{\delta}_{g}^{* (\ell)})  \\ & \cdot p(\boldsymbol{\delta}_{g}^{* (\ell+1)} \mid \kappa^{2 (\ell)}_{\zeta g}) \\
& \propto \exp \Big(-\frac{\omega_{ig}\eta_{ig}^2}{2} \Big) \cdot p(\omega_{ig}\mid 1,0)\vert\boldsymbol{B}_g\vert ^{-P/2}\exp\Big( -\frac{1}{2} \boldsymbol{\delta}_{g}^{*\prime}\boldsymbol{B}_g^{-1}\boldsymbol{\delta}_{g}^*\Big),
\end{aligned}
\end{equation*}
where $\eta_{ig}=\boldsymbol{V}_i^{*\prime}\boldsymbol{\delta}_{g}^*-C_{ig}$ and $C_{ig}=\log\sum_{h \neq j} \exp(\boldsymbol{V}_i^{*\prime}\boldsymbol{\delta}_{h}^*)$, $p(\omega_{ig}\mid 1,0)$ is the P\'{o}lya-gamma distribution $PG(b,c)$ with $b=1$ and $c=0$, $\boldsymbol{B}_g$ is the prior covariance matrix of Section 4.4 and $\boldsymbol{B}_g={\rm diag}(\sigma_{\delta g}^2\boldsymbol{1}_{P+1},\kappa^2_{\zeta g}\boldsymbol{1}_N)$. By assuming the conjugate prior $N(\boldsymbol{0},\boldsymbol{B}_g)$ on $\boldsymbol{\delta}_{g}^*$, the posterior distribution of the P\'{o}lya-gamma latent variable is
\begin{equation*}
(\omega_{ig}^{(\ell+1)} \mid \boldsymbol{V}^*, \boldsymbol{\delta}_{g}^{* (\ell)}) \sim PG(1,\eta_{ig}).
\end{equation*}
Thus, the conditional distributions of the logistic parameters (including the random intercepts) are
\begin{equation*}
    (\boldsymbol{\delta}_{g}^{* (\ell+1)} \mid \boldsymbol{V}^*, z_{ig}^{ (\ell)},\omega_{ig}^{(\ell+1)},\kappa^{2 (\ell)}_{\zeta g}) \sim N(\boldsymbol{M}_{g}, \boldsymbol{\Sigma}_{g}),
\end{equation*}
where $\boldsymbol{\Sigma}_{g}=(\boldsymbol{V}^{*\prime}\boldsymbol{\Omega}_g\boldsymbol{V}^*+ \boldsymbol{B}_g^{-1})^{-1}$,  $\boldsymbol{M}_{g}=\boldsymbol{\Sigma}_{g} \big[ \boldsymbol{V}^{*\prime}(\boldsymbol{\Omega}_g\boldsymbol{C}_g + \boldsymbol{\xi}_g)\big]$,  $\boldsymbol{\Omega}_g={\rm diag}(\omega_{1g},\cdots,\omega_{Ng})$, $\boldsymbol{C}_g=(C_{1g},\cdots,C_{Ng})^{\prime}$, and $\boldsymbol{\xi}_g=(\xi_{1g},\cdots,\xi_{Ng})^{\prime}$, with $\xi_{ig}=z_{ig}-\frac{1}{2}$. Thus, $\boldsymbol{\delta}_{g}^{* (\ell+1)}$ is drawn by first sampling $(\omega_{ig}^{(\ell+1)} \mid \boldsymbol{V}^*, \boldsymbol{\delta}_{g}^{* (\ell)})$ and then $(\boldsymbol{\delta}_{g}^{* (\ell+1)} \mid \boldsymbol{V}^*, z_{ig}^{ (\ell)},\omega_{ig}^{(\ell+1)},\kappa^{2 (\ell)}_{\zeta g})$.

\item \textbf{Sampling the variances of the random intercepts}
    
By analogy with sampling the error variances and sthe moothing parameters, we first draw $(a_{\kappa_g}^{(\ell+1)} \mid \kappa^{2 (\ell)}_{\zeta g}) \sim IG \Big(\frac{\nu_{\kappa}+1}{2},\frac{\nu_{\kappa}}{\kappa^2_{\zeta g}}+\frac{1}{A_{\kappa}^2}\Big)$. The conditional posterior distributions of the variances of the random intercepts are
\begin{equation*}
\begin{aligned}
p(\kappa^{2 (\ell+1)}_{\zeta g} \mid \boldsymbol{\zeta}_g^{(\ell+1)},a_{\kappa_g}^{(\ell+1)}) & \propto p(\boldsymbol{\zeta}_g^{(\ell+1)} \mid \kappa^{2 (\ell+1)}_{\zeta g}) \cdot p(a_{\kappa_g}^{(\ell+1)} \mid \kappa^{2 (\ell+1)}_{\zeta g}) \cdot p(\kappa^{2 (\ell+1)}_{\zeta g})  \\
& \propto (\kappa^2_{\zeta g})^{-\frac{N+\nu_{\kappa}}{2}+1} \exp{\Big[-\frac{1}{\kappa^2_{\zeta g}} \Big(\frac{\nu_{\kappa}}{a_{\kappa_g}}+\frac{\boldsymbol{\zeta}_g^{T}\boldsymbol{\zeta}_g}{2}\Big)\Big]},
\end{aligned}
\end{equation*}
which is $IG \Big(\frac{\nu_{\kappa}+N}{2}, \frac{\boldsymbol{\zeta}_g^{T}\boldsymbol{\zeta}_g}{2} + \frac{\nu_{\kappa}}{a_{\kappa_g}}\Big)$. The sampling scheme proceeds by first sampling $(a_{\kappa_g}^{(\ell+1)} \mid \kappa^{2 (\ell)}_{\zeta g})$ and then $(\kappa^{2 (\ell+1)}_{\zeta g} \mid \boldsymbol{\zeta}_g^{(\ell+1)},a_{\kappa_g}^{(\ell+1)})$.

\item \textbf{Computing the mixing weights}
    
After drawing the $\boldsymbol{\delta}_g^*$, the mixing weights $\pi_{ig}^{(\ell+1)}$ for each component, given the design matrix $V_i^*$, are computed by
\begin{equation*}
    p(\pi_{ig}^{(\ell+1)} \mid V_i^*, \boldsymbol{\delta}_g^{* (\ell+1)})=\frac{\exp (\boldsymbol{V}_i^{*T} \boldsymbol{\delta}_g^*)} {\sum_{h=1}^G \exp (\boldsymbol{V}_i^{*T} \boldsymbol{\delta}_h^*)}.
\end{equation*}

\item \textbf{Sampling the latent indicators}
    
After sampling all parameters and computing the mixing weights, the final Gibbs step is to allocate subjects to different components by drawing the latent indicators $z_{ig}$. As in Section 3.3, the conditional posterior of these indicators is  
\begin{equation*}
p(z_{ig}^{(\ell+1)}=1 \mid \boldsymbol{y}, \boldsymbol{S}, \boldsymbol{\Theta}^{(\ell+1)}, \pi_{ig}^{(\ell+1)})= \frac{\pi_{ig} \prod_{k=1}^K f_{gk}(\boldsymbol{y}_{ik} \mid \boldsymbol{\Theta}_{gk})}{\sum_{h=1}^G  \pi_{ih} \prod_{k=1}^K f_{hk}(\boldsymbol{y}_{ik} \mid \boldsymbol{\Theta}_{hk})},
\end{equation*}
and the indicators are drawn from the multinomial distribution.
    
\end{enumerate}

\subsection*{Appendix B: Additional simulation results}

Appendix B adds more simulation results in addition to simulation results in the paper itself. To further demonstrate the performance of the proposed method, we conduct simulation studies under two scenarios: two-component mixture
of trivariate time series and four-component mixture of bivariate time series. The model formula is displayed in Section 6.1 of the paper. We investigate the performance of our proposed method in terms of estimated trajectories and logistic parameters.

Mean(SD) of the ARSE, A-bias and V-bias for each component of the two-component trivariate model are given in Table \ref{Tablesim2gesttraj}. To demonstrate the performance of the proposed method in various settings, we look at combinations of the number of multivariate time series ($N=150, 250$) and the length of each time series ($n=50, 70$), and compare our proposed method to two existing methods: \texttt{gbmt} package in R \citep{magrini2022assessment} and \texttt{TRAJ} procedure in \texttt{SAS} \citep{nagin2018group}. The case of $n=50$ and $N=150$ in Table \ref{Tablesim2gesttraj} corresponds to Figure 3 in the main paper. The performance of the logistic parameters (RMSEs) with different values of $n$ and $N$ are given in Table 1 of the paper.

The Mean(SD) of the ARSE, A-bias and V-bias for each component of the $N=150$ four-component mixture of bivariate time series of length $n=50$ are given in Table \ref{Tablesim4gn50N150}, which corresponds to Figure 4 in the main paper. RMSEs of the logistic parameters for this setting are listed in Table 2 of the paper. Tables 5 - 10 present performance measures of the estimated trajectories and logistic parameters for combinations of different lengths of time series $n$ and numbers of time series $N$, under the scenario of the four-component bivariate model.

As expected, our proposed method outperforms the two existing methods in terms of the estimated trajectories for each component under different settings (different values of $n$ and $N$, for both the two-component trivariate and the four-component bivariate scenarios). The proposed method is able to achieve smaller ARSE and V-bias, while all three methods are able to obtain estimated trajectories with a very small bias. Notably, for the four-component bivariate scenario, \texttt{TRAJ} gives larger values of mean ARSE, A-bias and V-bias, which result from imprecise estimates of several replicates due to convergence issues. In terms of the logistic parameters, our proposed method outperforms \texttt{TRAJ} in almost all comparisons, especially for the intercept $\delta_0$ and the slope of the first covariate $\delta_1$. Our proposed method yields shrinkage estimates for the logistic parameters due to using a Bayesian method, while the multinomial logistic regression used in  \texttt{TRAJ}  gives inflated parameter estimates in case of perfect separations and unbalanced designs. 

\begin{table}[!p]
\caption{Mean (standard deviation) of the averaged root square error (ARSE), the averaged bias (A-bias) and the variance of bias (V-bias) of estimated trajectories for each component from $100$ replicates of $N$ two-component trivariate time series of length $n$. The proposed method was compared to \texttt{R} package \texttt{gbmt} and \texttt{TRAJ} procedure in \texttt{SAS}. C1 and C2 denote first and second components. Means were calculated by averaging over estimates of $100$ replicates. Standard deviations are Monte Carlo standard deviations from estimates of $100$ replicates. Each value was reported $\times 10^2$.
\label{Tablesim2gesttraj}}
\centering
{\tabcolsep=5pt
\begin{tabular}{@{}ccccccccc@{}}
\hline
n & N & Method & ARSE C1& A-bias C1 & V-bias C1 & ARSE C2 & A-bias C2 & V-bias C2 \\ \hline
                                         
\multirow{5}{*}{50} & \multirow{5}{*}{150} & Proposed & \begin{tabular}[c]{@{}c@{}}8.35\\ (1.26)\end{tabular}  & \begin{tabular}[c]{@{}c@{}}0.03\\ (1.83)\end{tabular} & \begin{tabular}[c]{@{}c@{}}0.68\\ (0.22)\end{tabular} & \begin{tabular}[c]{@{}c@{}}7.65\\ (1.38)\end{tabular}  & \begin{tabular}[c]{@{}c@{}}0.10\\ (1.76)\end{tabular}  & \begin{tabular}[c]{@{}c@{}}0.58\\ (0.22)\end{tabular} \\  
                    &                      & gbmt     & \begin{tabular}[c]{@{}c@{}}10.67\\ (1.91)\end{tabular} & \begin{tabular}[c]{@{}c@{}}0.03\\ (1.83)\end{tabular} & \begin{tabular}[c]{@{}c@{}}1.15\\ (0.42)\end{tabular} & \begin{tabular}[c]{@{}c@{}}9.08\\ (1.72)\end{tabular}  & \begin{tabular}[c]{@{}c@{}}0.10\\ (1.76)\end{tabular}  & \begin{tabular}[c]{@{}c@{}}0.83\\ (0.33)\end{tabular} \\ 
                    &                      & TRAJ     & \begin{tabular}[c]{@{}c@{}}11.06\\ (1.48)\end{tabular} & \begin{tabular}[c]{@{}c@{}}0.03\\ (1.83)\end{tabular} & \begin{tabular}[c]{@{}c@{}}1.22\\ (0.33)\end{tabular} & \begin{tabular}[c]{@{}c@{}}10.59\\ (1.52)\end{tabular} & \begin{tabular}[c]{@{}c@{}}0.10\\ (1.76)\end{tabular}  & \begin{tabular}[c]{@{}c@{}}1.12\\ (0.34)\end{tabular} \\ \hline
\multirow{5}{*}{70} & \multirow{5}{*}{150} & Proposed & \begin{tabular}[c]{@{}c@{}}7.16\\ (1.04)\end{tabular}  & \begin{tabular}[c]{@{}c@{}}0.24\\ (1.38)\end{tabular} & \begin{tabular}[c]{@{}c@{}}0.51\\ (0.16)\end{tabular} & \begin{tabular}[c]{@{}c@{}}6.53\\ (1.07)\end{tabular}  & \begin{tabular}[c]{@{}c@{}}-0.11\\ (1.42)\end{tabular} & \begin{tabular}[c]{@{}c@{}}0.42\\ (0.14)\end{tabular} \\ 
                    &                      & gbmt     & \begin{tabular}[c]{@{}c@{}}9.91\\ (1.96)\end{tabular}  & \begin{tabular}[c]{@{}c@{}}0.24\\ (1.38)\end{tabular} & \begin{tabular}[c]{@{}c@{}}1.01\\ (0.40)\end{tabular} & \begin{tabular}[c]{@{}c@{}}8.19\\ (1.65)\end{tabular}  & \begin{tabular}[c]{@{}c@{}}-0.11\\ (1.42)\end{tabular} & \begin{tabular}[c]{@{}c@{}}0.68\\ (0.29)\end{tabular} \\ 
                    &                      & TRAJ     & \begin{tabular}[c]{@{}c@{}}9.34\\ (1.10)\end{tabular}  & \begin{tabular}[c]{@{}c@{}}0.24\\ (1.38)\end{tabular} & \begin{tabular}[c]{@{}c@{}}0.87\\ (0.22)\end{tabular} & \begin{tabular}[c]{@{}c@{}}8.95\\ (1.13)\end{tabular}  & \begin{tabular}[c]{@{}c@{}}-0.11\\ (1.42)\end{tabular}  & \begin{tabular}[c]{@{}c@{}}0.80\\ (0.20)\end{tabular} \\ \hline
\multirow{5}{*}{50} & \multirow{5}{*}{250} & Proposed & \begin{tabular}[c]{@{}c@{}}6.81\\ (1.02)\end{tabular}  & \begin{tabular}[c]{@{}c@{}}0.07\\ (1.33)\end{tabular} & \begin{tabular}[c]{@{}c@{}}0.46\\ (0.14)\end{tabular} & \begin{tabular}[c]{@{}c@{}}6.22\\ (0.94)\end{tabular}  & \begin{tabular}[c]{@{}c@{}}0.02\\ (1.31)\end{tabular}  & \begin{tabular}[c]{@{}c@{}}0.38\\ (0.12)\end{tabular} \\  
                    &                      & gbmt     & \begin{tabular}[c]{@{}c@{}}9.79\\ (1.91)\end{tabular}  & \begin{tabular}[c]{@{}c@{}}0.07\\ (1.33)\end{tabular} & \begin{tabular}[c]{@{}c@{}}0.98\\ (0.40)\end{tabular} & \begin{tabular}[c]{@{}c@{}}8.00\\ (1.53)\end{tabular}  & \begin{tabular}[c]{@{}c@{}}0.02\\ (1.31)\end{tabular}  & \begin{tabular}[c]{@{}c@{}}0.65\\ (0.26)\end{tabular} \\  
                    &                      & TRAJ     & \begin{tabular}[c]{@{}c@{}}8.70\\ (1.18)\end{tabular} & \begin{tabular}[c]{@{}c@{}}0.07\\ (1.33)\end{tabular} & \begin{tabular}[c]{@{}c@{}}0.76\\ (0.21)\end{tabular} & \begin{tabular}[c]{@{}c@{}}8.20\\ (1.03)\end{tabular}  & \begin{tabular}[c]{@{}c@{}}0.02\\ (1.31)\end{tabular}  & \begin{tabular}[c]{@{}c@{}}0.67\\ (0.17)\end{tabular} \\ \hline
\multirow{5}{*}{70} & \multirow{5}{*}{250} & Proposed & \begin{tabular}[c]{@{}c@{}}5.65\\ (0.84)\end{tabular}  & \begin{tabular}[c]{@{}c@{}}0.08\\ (1.00)\end{tabular} & \begin{tabular}[c]{@{}c@{}}0.32\\ (0.10)\end{tabular} & \begin{tabular}[c]{@{}c@{}}5.27\\ (0.82)\end{tabular}  & \begin{tabular}[c]{@{}c@{}}-0.06\\ (1.42)\end{tabular} & \begin{tabular}[c]{@{}c@{}}0.27\\ (0.09)\end{tabular} \\ 
                    &                      & gbmt     & \begin{tabular}[c]{@{}c@{}}9.15\\ (1.96)\end{tabular}  & \begin{tabular}[c]{@{}c@{}}0.08\\ (1.00)\end{tabular} & \begin{tabular}[c]{@{}c@{}}0.87\\ (0.38)\end{tabular} & \begin{tabular}[c]{@{}c@{}}7.43\\ (1.60)\end{tabular}  & \begin{tabular}[c]{@{}c@{}}-0.06\\ (1.42)\end{tabular} & \begin{tabular}[c]{@{}c@{}}0.56\\ (0.26)\end{tabular} \\  
                    &                      & TRAJ     & \begin{tabular}[c]{@{}c@{}}7.18\\ (0.94)\end{tabular}  & \begin{tabular}[c]{@{}c@{}}0.08\\ (1.00)\end{tabular} & \begin{tabular}[c]{@{}c@{}}0.52\\ (0.14)\end{tabular} & \begin{tabular}[c]{@{}c@{}}6.80\\ (0.77)\end{tabular}  & \begin{tabular}[c]{@{}c@{}}-0.06\\ (1.42)\end{tabular} & \begin{tabular}[c]{@{}c@{}}0.45\\ (0.10)\end{tabular} \\ \hline
\end{tabular}}
\end{table}

\begin{table}[!p]
\caption{Mean (standard deviation) of the averaged root square error (ARSE), the averaged bias (A-bias) and the variance of bias (V-bias) of estimated trajectories for each component from $100$ replicates of $150$ four-component bivariate time series of length $50$. The proposed method was compared to \texttt{R} package \texttt{gbmt} and \texttt{TRAJ} procedure in \texttt{SAS}. C1, C2, C3 and C4 denote first, second, third and fourth component, respectively. Means were calculated by averaging over estimates of $100$ replicates. Standard deviations are Monte Carlo standard deviations from estimates of $100$ replicates. Each value was reported $\times 10^2$.
\label{Tablesim4gn50N150}}
\centering
{\tabcolsep=5pt
\begin{tabular}{@{}ccccccccc@{}}
\hline
n & N & Method & ARSE C1& A-bias C1 & V-bias C1 & ARSE C2 & A-bias C2 & V-bias C2 \\ \hline
                                         
\multirow{5}{*}{50} & \multirow{5}{*}{150} & Proposed & \begin{tabular}[c]{@{}c@{}}4.38\\ (1.04)\end{tabular}  & \begin{tabular}[c]{@{}c@{}}0.38\\ (1.59)\end{tabular} & \begin{tabular}[c]{@{}c@{}}0.18\\ (0.08)\end{tabular} & \begin{tabular}[c]{@{}c@{}}3.76\\ (0.87)\end{tabular}  & \begin{tabular}[c]{@{}c@{}}-0.01\\ (1.37)\end{tabular}  & \begin{tabular}[c]{@{}c@{}}0.13\\ (0.06)\end{tabular} \\  
                    &                      & gbmt     & \begin{tabular}[c]{@{}c@{}}4.75\\ (1.04)\end{tabular} & \begin{tabular}[c]{@{}c@{}}0.38\\ (1.59)\end{tabular} & \begin{tabular}[c]{@{}c@{}}0.21\\ (0.09)\end{tabular} & \begin{tabular}[c]{@{}c@{}}4.79\\ (1.17)\end{tabular}  & \begin{tabular}[c]{@{}c@{}}0.01\\ (1.65)\end{tabular}  & \begin{tabular}[c]{@{}c@{}}0.22\\ (0.11)\end{tabular} \\ 
                    &                      & TRAJ     & \begin{tabular}[c]{@{}c@{}}13.87\\ (15.59)\end{tabular} & \begin{tabular}[c]{@{}c@{}}0.62\\ (9.91)\end{tabular} & \begin{tabular}[c]{@{}c@{}}3.39\\ (9.24)\end{tabular} & \begin{tabular}[c]{@{}c@{}}12.41\\ (13.42)\end{tabular} & \begin{tabular}[c]{@{}c@{}}0.08\\ (9.74)\end{tabular}  & \begin{tabular}[c]{@{}c@{}}2.41\\ (5.37)\end{tabular} \\ \hline
                    
n & N & Method & ARSE C3& A-bias C3 & V-bias C3 & ARSE C4 & A-bias C4 & V-bias C4 \\ \hline

\multirow{5}{*}{50} & \multirow{5}{*}{150} & Proposed & \begin{tabular}[c]{@{}c@{}}4.69\\ (1.14)\end{tabular}  & \begin{tabular}[c]{@{}c@{}}-0.11\\ (1.83)\end{tabular} & \begin{tabular}[c]{@{}c@{}}0.20\\ (0.12)\end{tabular} & \begin{tabular}[c]{@{}c@{}}3.88\\ (1.15)\end{tabular}  & \begin{tabular}[c]{@{}c@{}}-0.09\\ (1.56)\end{tabular} & \begin{tabular}[c]{@{}c@{}}0.14\\ (0.08)\end{tabular} \\ 
                    &                      & gbmt     & \begin{tabular}[c]{@{}c@{}}5.08\\ (1.12)\end{tabular}  & \begin{tabular}[c]{@{}c@{}}-0.12\\ (1.83)\end{tabular} & \begin{tabular}[c]{@{}c@{}}0.24\\ (0.12)\end{tabular} & \begin{tabular}[c]{@{}c@{}}4.70\\ (1.32)\end{tabular}  & \begin{tabular}[c]{@{}c@{}}-0.09\\ (1.78)\end{tabular} & \begin{tabular}[c]{@{}c@{}}0.21\\ (0.12)\end{tabular} \\ 
                    &                      & TRAJ     & \begin{tabular}[c]{@{}c@{}}14.55\\ (14.82)\end{tabular}  & \begin{tabular}[c]{@{}c@{}}-1.58\\ (10.31)\end{tabular} & \begin{tabular}[c]{@{}c@{}}3.24\\ (7.35)\end{tabular} & \begin{tabular}[c]{@{}c@{}}14.36\\ (17.01)\end{tabular}  & \begin{tabular}[c]{@{}c@{}}0.12\\ (9.92)\end{tabular}  & \begin{tabular}[c]{@{}c@{}}3.99\\ (10.70)\end{tabular} \\ \hline

\end{tabular}}
\end{table}

\begin{table}[!p]
\caption{Mean (standard deviation) of the averaged root square error (ARSE), the averaged bias (A-bias) and the variance of bias (V-bias) of estimated trajectories for each component from $100$ replicates of $150$ four-component bivariate time series of length $70$. The proposed method was compared to \texttt{R} package \texttt{gbmt} and \texttt{TRAJ} procedure in \texttt{SAS}. C1, C2, C3 and C4 denote first, second, third and fourth component, respectively. Means were calculated by averaging over estimates of $100$ replicates. Standard deviations are Monte Carlo standard deviations from estimates of $100$ replicates. Each value was reported $\times 10^2$.
\label{Tablesim4gn70N150}}
\centering
{\tabcolsep=5pt
\begin{tabular}{@{}ccccccccc@{}}
\hline
n & N & Method & ARSE C1& A-bias C1 & V-bias C1 & ARSE C2 & A-bias C2 & V-bias C2 \\ \hline
                                         
\multirow{5}{*}{70} & \multirow{5}{*}{150} & Proposed & \begin{tabular}[c]{@{}c@{}}3.82\\ (0.95)\end{tabular}  & \begin{tabular}[c]{@{}c@{}}0.44\\ (1.30)\end{tabular} & \begin{tabular}[c]{@{}c@{}}0.14\\ (0.07)\end{tabular} & \begin{tabular}[c]{@{}c@{}}3.22\\ (0.85)\end{tabular}  & \begin{tabular}[c]{@{}c@{}}-0.07\\ (0.95)\end{tabular}  & \begin{tabular}[c]{@{}c@{}}0.10\\ (0.06)\end{tabular} \\  
                    &                      & gbmt     & \begin{tabular}[c]{@{}c@{}}4.05\\ (0.97)\end{tabular} & \begin{tabular}[c]{@{}c@{}}0.44\\ (1.30)\end{tabular} & \begin{tabular}[c]{@{}c@{}}0.16\\ (0.08)\end{tabular} & \begin{tabular}[c]{@{}c@{}}4.11\\ (1.12)\end{tabular}  & \begin{tabular}[c]{@{}c@{}}-0.08\\ (1.15)\end{tabular}  & \begin{tabular}[c]{@{}c@{}}0.17\\ (0.10)\end{tabular} \\ 
                    &                      & TRAJ     & \begin{tabular}[c]{@{}c@{}}13.51\\ (17.04)\end{tabular} & \begin{tabular}[c]{@{}c@{}}-0.30\\ (9.34)\end{tabular} & \begin{tabular}[c]{@{}c@{}}3.86\\ (10.73)\end{tabular} & \begin{tabular}[c]{@{}c@{}}10.00\\ (10.11)\end{tabular} & \begin{tabular}[c]{@{}c@{}}-0.25\\ (6.21)\end{tabular}  & \begin{tabular}[c]{@{}c@{}}1.64\\ (4.02)\end{tabular} \\ \hline
                    
n & N & Method & ARSE C3& A-bias C3 & V-bias C3 & ARSE C4 & A-bias C4 & V-bias C4 \\ \hline

\multirow{5}{*}{70} & \multirow{5}{*}{150} & Proposed & \begin{tabular}[c]{@{}c@{}}4.12\\ (0.90)\end{tabular}  & \begin{tabular}[c]{@{}c@{}}-0.29\\ (1.69)\end{tabular} & \begin{tabular}[c]{@{}c@{}}0.15\\ (0.06)\end{tabular} & \begin{tabular}[c]{@{}c@{}}3.52\\ (0.85)\end{tabular}  & \begin{tabular}[c]{@{}c@{}}0.24\\ (1.22)\end{tabular} & \begin{tabular}[c]{@{}c@{}}0.12\\ (0.06)\end{tabular} \\ 
                    &                      & gbmt     & \begin{tabular}[c]{@{}c@{}}4.38\\ (1.01)\end{tabular}  & \begin{tabular}[c]{@{}c@{}}-0.29\\ (1.69)\end{tabular} & \begin{tabular}[c]{@{}c@{}}0.17\\ (0.07)\end{tabular} & \begin{tabular}[c]{@{}c@{}}4.13\\ (0.99)\end{tabular}  & \begin{tabular}[c]{@{}c@{}}0.27\\ (1.40)\end{tabular} & \begin{tabular}[c]{@{}c@{}}0.16\\ (0.09)\end{tabular} \\ 
                    &                      & TRAJ     & \begin{tabular}[c]{@{}c@{}}13.03\\ (17.04)\end{tabular}  & \begin{tabular}[c]{@{}c@{}}0.30\\ (10.49)\end{tabular} & \begin{tabular}[c]{@{}c@{}}3.86\\ (10.73)\end{tabular} & \begin{tabular}[c]{@{}c@{}}11.77\\ (13.78)\end{tabular}  & \begin{tabular}[c]{@{}c@{}}0.39\\ (8.49)\end{tabular}  & \begin{tabular}[c]{@{}c@{}}2.57\\ (6.45)\end{tabular} \\ \hline

\end{tabular}}
\end{table}

\begin{table}[!p]
\caption{Root mean square errors (RMSEs) of each logistic parameter for the four-component bivariate model from $100$ replicates of $150$ four-component bivariate time series of length $70$. RMSEs of the proposed method were compared to \texttt{TRAJ} procedure in \texttt{SAS}. Parameters $\delta_0$, $\delta_1$, $\delta_2$ and $\delta_3$ are intercept, first, second and third logistic parameters, respectively. The fourth component was used as the reference component. The true values of logistic parameters are $5,-3.5,1,0.1$ (first component), $-4,2.5,-2,-0.2$ (second component), $3,-2,0.8,0.2$ (third component). C1, C2, C3 and C4 denote first, second, third and fourth component, respectively.
\label{Tablelog4gn70N150}}
\centering
{\tabcolsep=12pt
\begin{tabular}{@{}clcccccc@{}}
\hline
\multicolumn{1}{l}{n} & N & Method & Comparison & $\delta_0$ & $\delta_1$ & $\delta_2$ & $\delta_3$ \\ \hline

\multirow{6}{*}{70} & \multirow{6}{*}{150} & \multirow{3}{*}{Proposed}      & C1 vs C4   & 0.81  & 0.51 & 0.29 & 0.41 \\ 
                             & & & C2 vs C4   & 1.42 & 0.73 & 0.58 & 0.36 \\  
                             & & & C3 vs C4   & 1.05 & 0.58 & 0.37 & 0.31 \\ \cline{3-8}
& & \multirow{3}{*}{TRAJ}   & C1 vs C4   & 1.13 & 0.66 & 0.31 & 0.45 \\  
                             & & & C2 vs C4   & 3.12 & 1.66 & 0.99 & 0.55 \\ 
                             & & & C3 vs C4   & 1.15 & 0.74 & 0.48 & 0.35 \\ \hline
\end{tabular}
}
\end{table}

\begin{table}[!p]
\caption{Mean (standard deviation) of the averaged root square error (ARSE), the averaged bias (A-bias) and the variance of bias (V-bias) of estimated trajectories for each component from $100$ replicates of $250$ four-component bivariate time series of length $50$. The proposed method was compared to \texttt{R} package \texttt{gbmt} and \texttt{TRAJ} procedure in \texttt{SAS}. C1, C2, C3 and C4 denote first, second, third and fourth component, respectively. Means were calculated by averaging over estimates of $100$ replicates. Standard deviations are Monte Carlo standard deviations from estimates of $100$ replicates. Each value was reported $\times 10^2$.
\label{Tablesim4gn50N250}}
\centering
{\tabcolsep=5pt
\begin{tabular}{@{}ccccccccc@{}}
\hline
n & N & Method & ARSE C1& A-bias C1 & V-bias C1 & ARSE C2 & A-bias C2 & V-bias C2 \\ \hline
                                         
\multirow{5}{*}{50} & \multirow{5}{*}{250} & Proposed & \begin{tabular}[c]{@{}c@{}}3.42\\ (0.78)\end{tabular}  & \begin{tabular}[c]{@{}c@{}}0.18\\ (1.19)\end{tabular} & \begin{tabular}[c]{@{}c@{}}0.11\\ (0.05)\end{tabular} & \begin{tabular}[c]{@{}c@{}}2.86\\ (0.61)\end{tabular}  & \begin{tabular}[c]{@{}c@{}}-0.14\\ (0.98)\end{tabular}  & \begin{tabular}[c]{@{}c@{}}0.08\\ (0.04)\end{tabular} \\  
                    &                      & gbmt     & \begin{tabular}[c]{@{}c@{}}3.57\\ (0.85)\end{tabular} & \begin{tabular}[c]{@{}c@{}}0.18\\ (1.19)\end{tabular} & \begin{tabular}[c]{@{}c@{}}0.12\\ (0.06)\end{tabular} & \begin{tabular}[c]{@{}c@{}}3.68\\ (0.79)\end{tabular}  & \begin{tabular}[c]{@{}c@{}}-0.15\\ (1.20)\end{tabular}  & \begin{tabular}[c]{@{}c@{}}0.13\\ (0.06)\end{tabular} \\ 
                    &                      & TRAJ     & \begin{tabular}[c]{@{}c@{}}11.66\\ (15.03)\end{tabular} & \begin{tabular}[c]{@{}c@{}}0.53\\ (10.63)\end{tabular} & \begin{tabular}[c]{@{}c@{}}2.50\\ (6.30)\end{tabular} & \begin{tabular}[c]{@{}c@{}}8.90\\ (9.38)\end{tabular} & \begin{tabular}[c]{@{}c@{}}1.47\\ (7.81)\end{tabular}  & \begin{tabular}[c]{@{}c@{}}1.05\\ (2.16)\end{tabular} \\ \hline
                    
n & N & Method & ARSE C3& A-bias C3 & V-bias C3 & ARSE C4 & A-bias C4 & V-bias C4 \\ \hline

\multirow{5}{*}{50} & \multirow{5}{*}{250} & Proposed & \begin{tabular}[c]{@{}c@{}}3.93\\ (0.92)\end{tabular}  & \begin{tabular}[c]{@{}c@{}}-0.06\\ (1.48)\end{tabular} & \begin{tabular}[c]{@{}c@{}}0.14\\ (0.07)\end{tabular} & \begin{tabular}[c]{@{}c@{}}3.28\\ (0.76)\end{tabular}  & \begin{tabular}[c]{@{}c@{}}-0.10\\ (1.19)\end{tabular} & \begin{tabular}[c]{@{}c@{}}0.10\\ (0.05)\end{tabular} \\ 
                    &                      & gbmt     & \begin{tabular}[c]{@{}c@{}}4.16\\ (0.95)\end{tabular}  & \begin{tabular}[c]{@{}c@{}}-0.06\\ (1.49)\end{tabular} & \begin{tabular}[c]{@{}c@{}}0.16\\ (0.07)\end{tabular} & \begin{tabular}[c]{@{}c@{}}3.83\\ (0.83)\end{tabular}  & \begin{tabular}[c]{@{}c@{}}-0.13\\ (1.36)\end{tabular} & \begin{tabular}[c]{@{}c@{}}0.14\\ (0.06)\end{tabular} \\ 
                    &                      & TRAJ     & \begin{tabular}[c]{@{}c@{}}10.80\\ (9.92)\end{tabular}  & \begin{tabular}[c]{@{}c@{}}0.49\\ (5.78)\end{tabular} & \begin{tabular}[c]{@{}c@{}}1.83\\ (3.70)\end{tabular} & \begin{tabular}[c]{@{}c@{}}10.17\\ (12.54)\end{tabular}  & \begin{tabular}[c]{@{}c@{}}-0.17\\ (8.83)\end{tabular}  & \begin{tabular}[c]{@{}c@{}}1.84\\ (5.20)\end{tabular} \\ \hline

\end{tabular}}
\end{table}

\begin{table}[!p]
\caption{Root mean square errors (RMSEs) of each logistic parameter for the four-component bivariate model from $100$ replicates of $250$ four-component bivariate time series of length $50$. RMSEs of the proposed method were compared to \texttt{TRAJ} procedure in \texttt{SAS}. Parameters $\delta_0$, $\delta_1$, $\delta_2$ and $\delta_3$ are intercept, first, second and third logistic parameters, respectively. The fourth component was used as the reference component. The true values of logistic parameters are $5,-3.5,1,0.1$ (first component), $-4,2.5,-2,-0.2$ (second component), $3,-2,0.8,0.2$ (third component). C1, C2, C3 and C4 denote first, second, third and fourth component, respectively.
\label{Tablelog4gn50N250}}
\centering
{\tabcolsep=12pt
\begin{tabular}{@{}clcccccc@{}}
\hline
\multicolumn{1}{l}{n} & N & Method & Comparison & $\delta_0$ & $\delta_1$ & $\delta_2$ & $\delta_3$ \\ \hline

\multirow{6}{*}{50} & \multirow{6}{*}{250} & \multirow{3}{*}{Proposed}      & C1 vs C4   & 0.63  & 0.41 & 0.26 & 0.29 \\ 
                             & & & C2 vs C4   & 1.00 & 0.46 & 0.40 & 0.27 \\  
                             & & & C3 vs C4   & 0.63 & 0.33 & 0.23 & 0.24 \\ \cline{3-8}
& & \multirow{3}{*}{TRAJ}   & C1 vs C4   & 0.91 & 0.56 & 0.30 & 0.28 \\  
                             & & & C2 vs C4   & 1.40 & 0.86 & 0.61 & 0.35 \\ 
                             & & & C3 vs C4   & 2.24 & 1.40 & 0.85 & 0.27 \\ \hline
\end{tabular}
}
\end{table}

\begin{table}[!p]
\caption{Mean (standard deviation) of the averaged root square error (ARSE), the averaged bias (A-bias) and the variance of bias (V-bias) of estimated trajectories for each component from $100$ replicates of $250$ four-component bivariate time series of length $70$. The proposed method was compared to \texttt{R} package \texttt{gbmt} and \texttt{TRAJ} procedure in \texttt{SAS}. C1, C2, C3 and C4 denote first, second, third and fourth component, respectively. Means were calculated by averaging over estimates of $100$ replicates. Standard deviations are Monte Carlo standard deviations from estimates of $100$ replicates. Each value was reported $\times 10^2$.
\label{Tablesim4gn70N250}}
\centering
{\tabcolsep=5pt
\begin{tabular}{@{}ccccccccc@{}}
\hline
n & N & Method & ARSE C1& A-bias C1 & V-bias C1 & ARSE C2 & A-bias C2 & V-bias C2 \\ \hline
                                         
\multirow{5}{*}{70} & \multirow{5}{*}{250} & Proposed & \begin{tabular}[c]{@{}c@{}}2.94\\ (0.60)\end{tabular}  & \begin{tabular}[c]{@{}c@{}}-0.04\\ (1.06)\end{tabular} & \begin{tabular}[c]{@{}c@{}}0.08\\ (0.04)\end{tabular} & \begin{tabular}[c]{@{}c@{}}2.61\\ (0.57)\end{tabular}  & \begin{tabular}[c]{@{}c@{}}-0.01\\ (0.87)\end{tabular}  & \begin{tabular}[c]{@{}c@{}}0.06\\ (0.03)\end{tabular} \\  
                    &                      & gbmt     & \begin{tabular}[c]{@{}c@{}}3.10\\ (0.63)\end{tabular} & \begin{tabular}[c]{@{}c@{}}-0.04\\ (1.06)\end{tabular} & \begin{tabular}[c]{@{}c@{}}0.09\\ (0.04)\end{tabular} & \begin{tabular}[c]{@{}c@{}}3.18\\ (0.70)\end{tabular}  & \begin{tabular}[c]{@{}c@{}}0.01\\ (1.05)\end{tabular}  & \begin{tabular}[c]{@{}c@{}}0.10\\ (0.05)\end{tabular} \\ 
                    &                      & TRAJ     & \begin{tabular}[c]{@{}c@{}}13.52\\ (17.70)\end{tabular} & \begin{tabular}[c]{@{}c@{}}-1.58\\ (11.09)\end{tabular} & \begin{tabular}[c]{@{}c@{}}3.71\\ (8.80)\end{tabular} & \begin{tabular}[c]{@{}c@{}}11.51\\ (14.85)\end{tabular} & \begin{tabular}[c]{@{}c@{}}-0.19\\ (9.98)\end{tabular}  & \begin{tabular}[c]{@{}c@{}}2.54\\ (7.10)\end{tabular} \\ \hline
                    
n & N & Method & ARSE C3& A-bias C3 & V-bias C3 & ARSE C4 & A-bias C4 & V-bias C4 \\ \hline

\multirow{5}{*}{70} & \multirow{5}{*}{250} & Proposed & \begin{tabular}[c]{@{}c@{}}3.30\\ (0.76)\end{tabular}  & \begin{tabular}[c]{@{}c@{}}-0.02\\ (1.21)\end{tabular} & \begin{tabular}[c]{@{}c@{}}0.10\\ (0.05)\end{tabular} & \begin{tabular}[c]{@{}c@{}}2.85\\ (0.73)\end{tabular}  & \begin{tabular}[c]{@{}c@{}}-0.07\\ (0.97)\end{tabular} & \begin{tabular}[c]{@{}c@{}}0.08\\ (0.04)\end{tabular} \\ 
                    &                      & gbmt     & \begin{tabular}[c]{@{}c@{}}3.51\\ (0.79)\end{tabular}  & \begin{tabular}[c]{@{}c@{}}-0.01\\ (1.21)\end{tabular} & \begin{tabular}[c]{@{}c@{}}0.12\\ (0.06)\end{tabular} & \begin{tabular}[c]{@{}c@{}}3.26\\ (0.80)\end{tabular}  & \begin{tabular}[c]{@{}c@{}}-0.09\\ (1.06)\end{tabular} & \begin{tabular}[c]{@{}c@{}}0.10\\ (0.05)\end{tabular} \\ 
                    &                      & TRAJ     & \begin{tabular}[c]{@{}c@{}}13.07\\ (15.21)\end{tabular}  & \begin{tabular}[c]{@{}c@{}}1.48\\ (10.52)\end{tabular} & \begin{tabular}[c]{@{}c@{}}2.90\\ (7.11)\end{tabular} & \begin{tabular}[c]{@{}c@{}}10.68\\ (12.93)\end{tabular}  & \begin{tabular}[c]{@{}c@{}}0.65\\ (8.11)\end{tabular}  & \begin{tabular}[c]{@{}c@{}}2.16\\ (5.66)\end{tabular} \\ \hline

\end{tabular}}
\end{table}

\begin{table}[!p]
\caption{Root mean square errors (RMSEs) of each logistic parameter for the four-component bivariate model from $100$ replicates of $250$ four-component bivariate time series of length $70$. RMSEs of the proposed method were compared to \texttt{TRAJ} procedure in \texttt{SAS}. Parameters $\delta_0$, $\delta_1$, $\delta_2$ and $\delta_3$ are intercept, first, second and third logistic parameters, respectively. The fourth component was used as the reference component. The true values of logistic parameters are $5,-3.5,1,0.1$ (first component), $-4,2.5,-2,-0.2$ (second component), $3,-2,0.8,0.2$ (third component). C1, C2, C3 and C4 denote first, second, third and fourth component, respectively.
\label{Tablelog4gn70N250}}
\centering
{\tabcolsep=12pt
\begin{tabular}{@{}clcccccc@{}}
\hline
\multicolumn{1}{l}{n} & N & Method & Comparison & $\delta_0$ & $\delta_1$ & $\delta_2$ & $\delta_3$ \\ \hline

\multirow{6}{*}{70} & \multirow{6}{*}{250} & \multirow{3}{*}{Proposed}      & C1 vs C4   & 0.64  & 0.40 & 0.26 & 0.28 \\ 
                             & & & C2 vs C4   & 0.92 & 0.42 & 0.41 & 0.28 \\  
                             & & & C3 vs C4   & 0.63 & 0.31 & 0.23 & 0.23 \\ \cline{3-8}
& & \multirow{3}{*}{TRAJ}   & C1 vs C4   & 0.82 & 0.50 & 0.27 & 0.28 \\  
                             & & & C2 vs C4   & 1.47 & 0.86 & 0.61 & 0.36 \\ 
                             & & & C3 vs C4   & 1.60 & 0.96 & 0.57 & 0.25 \\ \hline
\end{tabular}
}
\end{table}

\subsection*{Appendix C: Additional real-data results}

\begin{figure}[!p]		
\centering\includegraphics[width=1.0\textwidth]{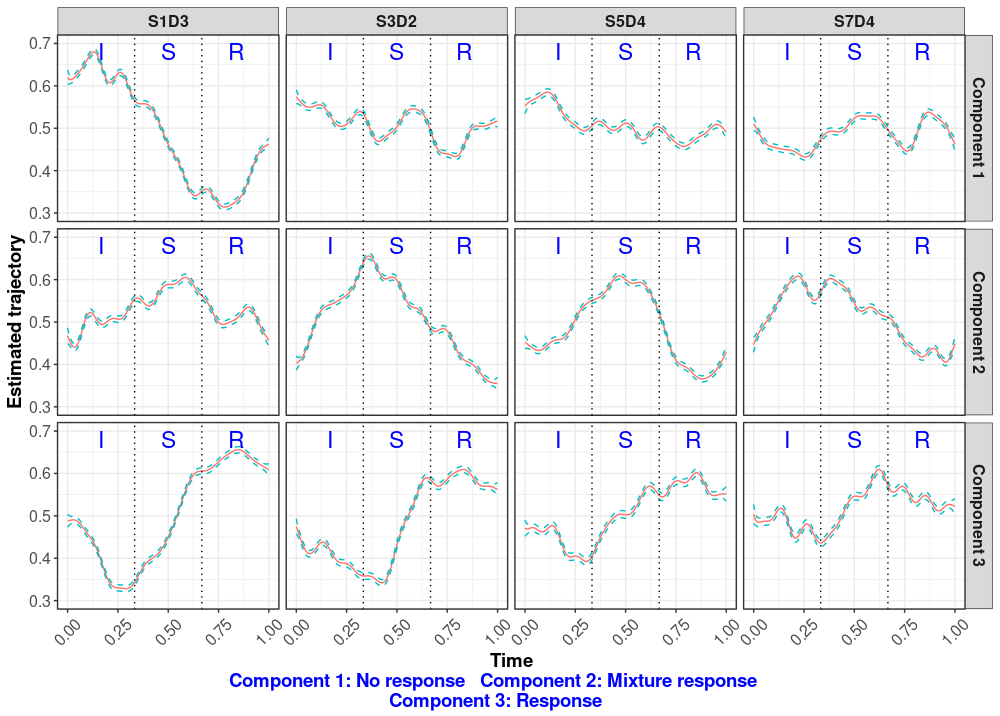}
\caption{Estimated trajectories of the three-component model with four selected channels. \textbf{I}: Interact \textbf{S}: Still-face \textbf{R}: Recovery. Red curves are posterior means and the two green dashed curves are 95\% pointwise credible intervals.}
\label{fig:fNIRS4g1357}
\end{figure}

\begin{figure}[!p]		
\centering\includegraphics[width=1.0\textwidth]{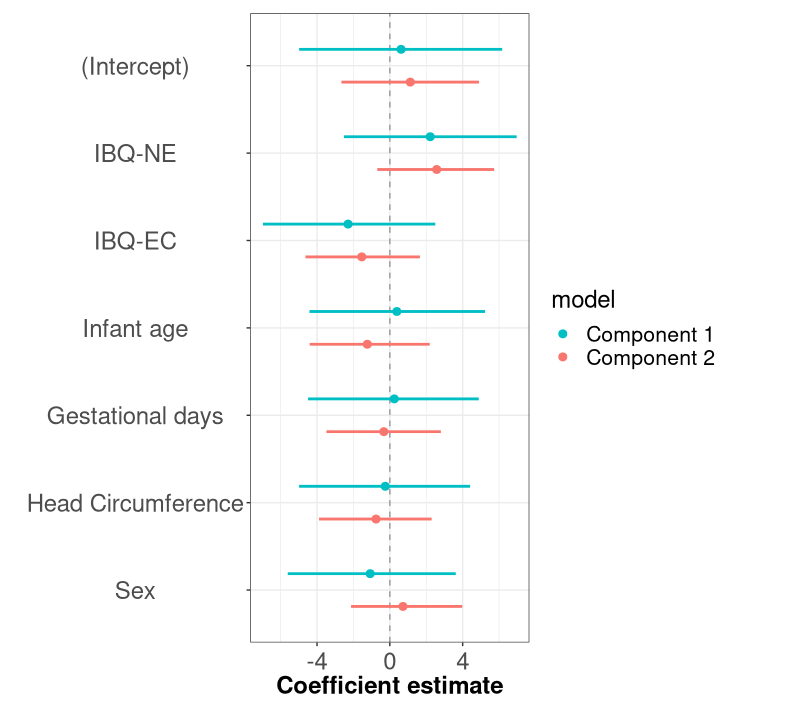}
\caption{Logistic coefficient estimates and 95\% credible intervals corresponding to each covariate of the three-component model.}
\label{fig:fNIRS4g1357log}
\end{figure}

Appendix C describes more real-data results in addition to those in Section 7 of the main paper. Our motivating study is described in Section 2 of the paper. Figure \ref{fig:fNIRS4g1357} shows the estimated trajectories of the three-component model for another set of four channels (S1D3, S3D2, S5D4, and S7D4). Based on the selection criterion DIC introduced in Section 5.2 of the main paper, the three-component model was selected as the best model. We named the second component as the mixture response component because it involves both increased and decreased brain activity or hemoglobin level for the still-face period for different channels. In addition, Figure \ref{fig:fNIRS4g1357log} displays the logistic coefficient estimates and 95\% credible intervals corresponding to each covariate. The last component (third component) is always used as the reference. We reach the same conclusion with positive estimates of IBQ-NE scores and negative estimates of IBQ-EC scores for both components (component 1 vs. 3, component 2 vs. 3).

\begin{figure}[!p]		
\centering\includegraphics[width=0.9\textwidth]{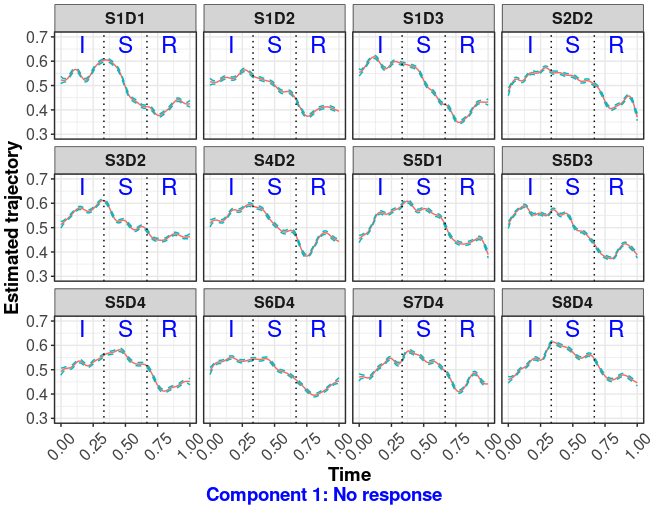}
\caption{Estimated trajectories of the first component for the three-component model with all twelve channels. \textbf{I}: Interact \textbf{S}: Still-face \textbf{R}: Recovery. Red curves are posterior mean and two green dashed curves are 95\% pointwise credible intervals.}
\label{fig:fNIRS12g1}
\end{figure}

\begin{figure}[!p]		
\centering\includegraphics[width=0.9\textwidth]{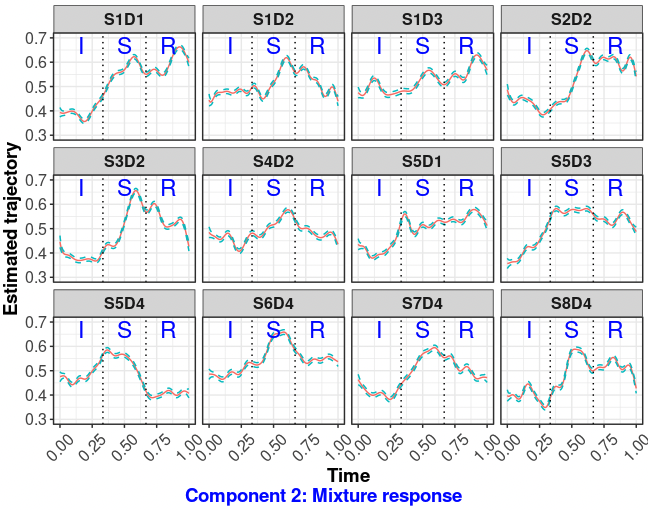}
\caption{Estimated trajectories of the second component for the three-component model with all twelve channels. \textbf{I}: Interact \textbf{S}: Still-face \textbf{R}: Recovery. Red curves are posterior mean and two green dashed curves are 95\% pointwise credible intervals.}
\label{fig:fNIRS12g2}
\end{figure}

\begin{figure}[!p]		
\centering\includegraphics[width=0.9\textwidth]{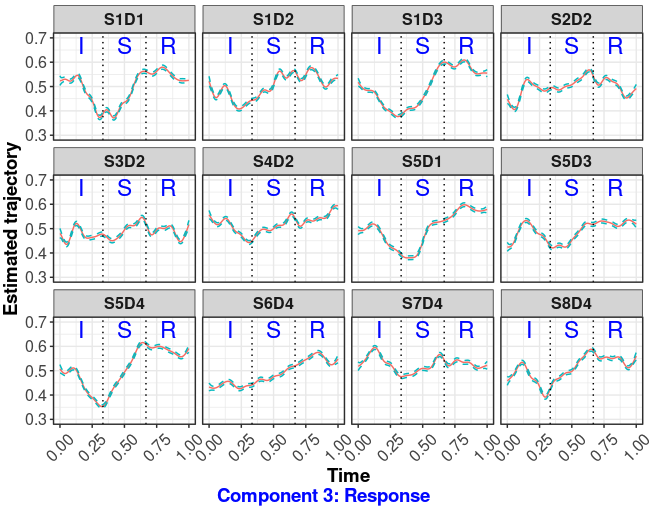}
\caption{Estimated trajectories of the third component for the three-component model with all twelve channels. \textbf{I}: Interact \textbf{S}: Still-face \textbf{R}: Recovery. Red curves are posterior mean and two green dashed curves are 95\% pointwise credible intervals.}
\label{fig:fNIRS12g3}
\end{figure}

\begin{figure}[!p]		
\centering\includegraphics[width=0.8\textwidth]{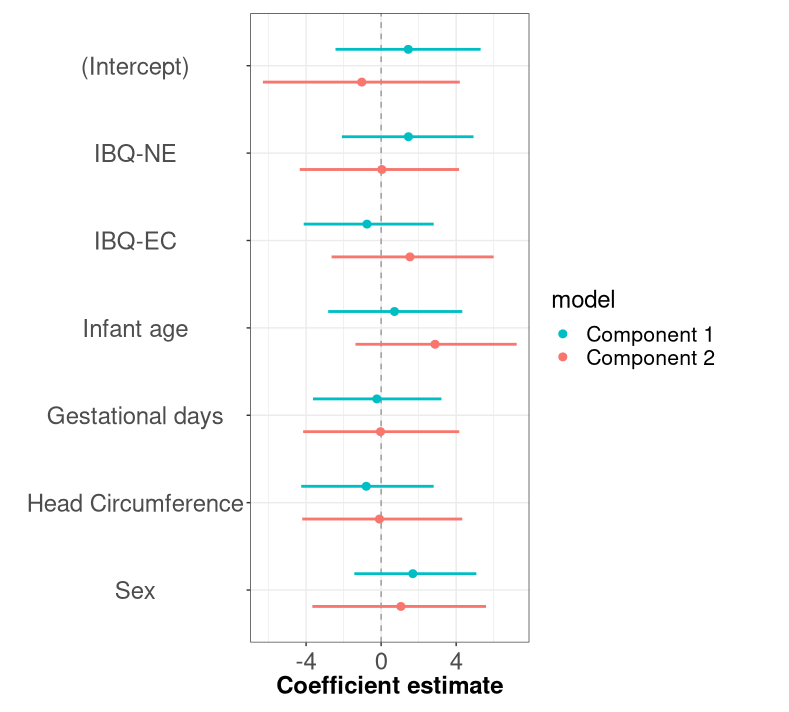}
\caption{Logistic coefficient estimates and 95\% credible intervals for each covariate of the three-component model for all twelve channels.}
\label{fig:fNIRS12glog}
\end{figure}

In addition to the four-channel analyses, we also present results from all channels (twelve channels). Figures \ref{fig:fNIRS12g1}, \ref{fig:fNIRS12g2}, \ref{fig:fNIRS12g3} present the estimated trajectories of the first, second and third component for the three-component model with all twelve channels, respectively. The three-component model was selected as the best model for the twelve-channel analysis based on the adjusted DIC. We named the three components no response, mixture response, and response component, respectively. Figure \ref{fig:fNIRS12glog}
displays the logistic coefficient estimates and 95 \% credible intervals corresponding to each covariate. 

\end{document}